\pdfoutput=1
\documentclass[11pt,twoside,a4paper,cmspaper,final,collab]{cms-tdr}
\def\svnVersion{2470}

\begin{document}\cmsNoteHeader{QCD-09-010}
%
%
%

%
%
\hyphenation{env-iron-men-tal}
\hyphenation{had-ron-i-za-tion}
\hyphenation{cal-or-i-me-ter}
\hyphenation{de-vices}
%
\RCS$Revision: 22 $
\RCS$HeadURL: svn+ssh://svn.cern.ch/reps/tdr2/papers/XXX-08-000/trunk/XXX-08-000.tex $
\RCS$Id: XXX-08-000.tex 22 2009-10-27 11:57:42Z alverson $
%
%
%

\newcommand {\etal}{\mbox{et al.}\xspace} 
\newcommand {\ie}{\mbox{i.e.}\xspace}     
\newcommand {\eg}{\mbox{e.g.}\xspace}     
\newcommand {\etc}{\mbox{etc.}\xspace}     
\newcommand {\vs}{\mbox{\sl vs.}\xspace}      
\newcommand {\mdash}{\ensuremath{\mathrm{-}}} 

\newcommand {\Lone}{Level-1\xspace} 
\newcommand {\Ltwo}{Level-2\xspace}
\newcommand {\Lthree}{Level-3\xspace}

\providecommand{\ACERMC} {\textsc{AcerMC}\xspace}
\providecommand{\ALPGEN} {{\textsc{alpgen}}\xspace}
\providecommand{\CHARYBDIS} {{\textsc{charybdis}}\xspace}
\providecommand{\CMKIN} {\textsc{cmkin}\xspace}
\providecommand{\CMSIM} {{\textsc{cmsim}}\xspace}
\providecommand{\CMSSW} {{\textsc{cmssw}}\xspace}
\providecommand{\COBRA} {{\textsc{cobra}}\xspace}
\providecommand{\COCOA} {{\textsc{cocoa}}\xspace}
\providecommand{\COMPHEP} {\textsc{CompHEP}\xspace}
\providecommand{\EVTGEN} {{\textsc{evtgen}}\xspace}
\providecommand{\FAMOS} {{\textsc{famos}}\xspace}
\providecommand{\GARCON} {\textsc{garcon}\xspace}
\providecommand{\GARFIELD} {{\textsc{garfield}}\xspace}
\providecommand{\GEANE} {{\textsc{geane}}\xspace}
\providecommand{\GEANTfour} {{\textsc{geant4}}\xspace}
\providecommand{\GEANTthree} {{\textsc{geant3}}\xspace}
\providecommand{\GEANT} {{\textsc{geant}}\xspace}
\providecommand{\HDECAY} {\textsc{hdecay}\xspace}
\providecommand{\HERWIG} {{\textsc{herwig}}\xspace}
\providecommand{\HIGLU} {{\textsc{higlu}}\xspace}
\providecommand{\HIJING} {{\textsc{hijing}}\xspace}
\providecommand{\IGUANA} {\textsc{iguana}\xspace}
\providecommand{\ISAJET} {{\textsc{isajet}}\xspace}
\providecommand{\ISAPYTHIA} {{\textsc{isapythia}}\xspace}
\providecommand{\ISASUGRA} {{\textsc{isasugra}}\xspace}
\providecommand{\ISASUSY} {{\textsc{isasusy}}\xspace}
\providecommand{\ISAWIG} {{\textsc{isawig}}\xspace}
\providecommand{\MADGRAPH} {\textsc{MadGraph}\xspace}
\providecommand{\MCATNLO} {\textsc{mc@nlo}\xspace}
\providecommand{\MCFM} {\textsc{mcfm}\xspace}
\providecommand{\MILLEPEDE} {{\textsc{millepede}}\xspace}
\providecommand{\ORCA} {{\textsc{orca}}\xspace}
\providecommand{\OSCAR} {{\textsc{oscar}}\xspace}
\providecommand{\PHOTOS} {\textsc{photos}\xspace}
\providecommand{\PROSPINO} {\textsc{prospino}\xspace}
\providecommand{\PYTHIA} {{\textsc{pythia}}\xspace}
\providecommand{\SHERPA} {{\textsc{sherpa}}\xspace}
\providecommand{\TAUOLA} {\textsc{tauola}\xspace}
\providecommand{\TOPREX} {\textsc{TopReX}\xspace}
\providecommand{\XDAQ} {{\textsc{xdaq}}\xspace}

\newcommand {\DZERO}{D\O\xspace}     


\newcommand{\de}{\ensuremath{^\circ}}
\newcommand{\ten}[1]{\ensuremath{\times \text{10}^\text{#1}}}
\newcommand{\unit}[1]{\ensuremath{\text{\,#1}}\xspace}
\newcommand{\mum}{\ensuremath{\,\mu\text{m}}\xspace}
\newcommand{\micron}{\ensuremath{\,\mu\text{m}}\xspace}
\newcommand{\cm}{\ensuremath{\,\text{cm}}\xspace}
\newcommand{\mm}{\ensuremath{\,\text{mm}}\xspace}
\newcommand{\mus}{\ensuremath{\,\mu\text{s}}\xspace}
\newcommand{\keV}{\ensuremath{\,\text{ke\hspace{-.08em}V}}\xspace}
\newcommand{\MeV}{\ensuremath{\,\text{Me\hspace{-.08em}V}}\xspace}
\newcommand{\GeV}{\ensuremath{\,\text{Ge\hspace{-.08em}V}}\xspace}
\newcommand{\TeV}{\ensuremath{\,\text{Te\hspace{-.08em}V}}\xspace}
\newcommand{\PeV}{\ensuremath{\,\text{Pe\hspace{-.08em}V}}\xspace}
\newcommand{\keVc}{\ensuremath{{\,\text{ke\hspace{-.08em}V\hspace{-0.16em}/\hspace{-0.08em}}c}}\xspace}
\newcommand{\MeVc}{\ensuremath{{\,\text{Me\hspace{-.08em}V\hspace{-0.16em}/\hspace{-0.08em}}c}}\xspace}
\newcommand{\GeVc}{\ensuremath{{\,\text{Ge\hspace{-.08em}V\hspace{-0.16em}/\hspace{-0.08em}}c}}\xspace}
\newcommand{\TeVc}{\ensuremath{{\,\text{Te\hspace{-.08em}V\hspace{-0.16em}/\hspace{-0.08em}}c}}\xspace}
\newcommand{\keVcc}{\ensuremath{{\,\text{ke\hspace{-.08em}V\hspace{-0.16em}/\hspace{-0.08em}}c^\text{2}}}\xspace}
\newcommand{\MeVcc}{\ensuremath{{\,\text{Me\hspace{-.08em}V\hspace{-0.16em}/\hspace{-0.08em}}c^\text{2}}}\xspace}
\newcommand{\GeVcc}{\ensuremath{{\,\text{Ge\hspace{-.08em}V\hspace{-0.16em}/\hspace{-0.08em}}c^\text{2}}}\xspace}
\newcommand{\TeVcc}{\ensuremath{{\,\text{Te\hspace{-.08em}V\hspace{-0.16em}/\hspace{-0.08em}}c^\text{2}}}\xspace}

\newcommand{\pbinv} {\mbox{\ensuremath{\,\text{pb}^\text{$-$1}}}\xspace}
\newcommand{\fbinv} {\mbox{\ensuremath{\,\text{fb}^\text{$-$1}}}\xspace}
\newcommand{\nbinv} {\mbox{\ensuremath{\,\text{nb}^\text{$-$1}}}\xspace}
\newcommand{\percms}{\ensuremath{\,\text{cm}^\text{$-$2}\,\text{s}^\text{$-$1}}\xspace}
\newcommand{\lumi}{\ensuremath{\mathcal{L}}\xspace}
\newcommand{\Lumi}{\ensuremath{\mathcal{L}}\xspace}
%
\newcommand{\LvLow}  {\ensuremath{\mathcal{L}=\text{10}^\text{32}\,\text{cm}^\text{$-$2}\,\text{s}^\text{$-$1}}\xspace}
\newcommand{\LLow}   {\ensuremath{\mathcal{L}=\text{10}^\text{33}\,\text{cm}^\text{$-$2}\,\text{s}^\text{$-$1}}\xspace}
\newcommand{\lowlumi}{\ensuremath{\mathcal{L}=\text{2}\times \text{10}^\text{33}\,\text{cm}^\text{$-$2}\,\text{s}^\text{$-$1}}\xspace}
\newcommand{\LMed}   {\ensuremath{\mathcal{L}=\text{2}\times \text{10}^\text{33}\,\text{cm}^\text{$-$2}\,\text{s}^\text{$-$1}}\xspace}
\newcommand{\LHigh}  {\ensuremath{\mathcal{L}=\text{10}^\text{34}\,\text{cm}^\text{-2}\,\text{s}^\text{$-$1}}\xspace}
\newcommand{\hilumi} {\ensuremath{\mathcal{L}=\text{10}^\text{34}\,\text{cm}^\text{-2}\,\text{s}^\text{$-$1}}\xspace}


\newcommand{\zp}{\ensuremath{\mathrm{Z}^\prime}\xspace}


\newcommand{\kt}{\ensuremath{k_{\mathrm{T}}}\xspace}
\newcommand{\BC}{\ensuremath{{B_{\mathrm{c}}}}\xspace}
\newcommand{\bbarc}{\ensuremath{{\overline{b}c}}\xspace}
\newcommand{\bbbar}{\ensuremath{{b\overline{b}}}\xspace}
\newcommand{\ccbar}{\ensuremath{{c\overline{c}}}\xspace}
\newcommand{\JPsi}{\ensuremath{{J}\hspace{-.08em}/\hspace{-.14em}\psi}\xspace}
\newcommand{\bspsiphi}{\ensuremath{B_s \to \JPsi\, \phi}\xspace}
\newcommand{\AFB}{\ensuremath{A_\text{FB}}\xspace}
\newcommand{\EE}{\ensuremath{e^+e^-}\xspace}
\newcommand{\MM}{\ensuremath{\mu^+\mu^-}\xspace}
\newcommand{\TT}{\ensuremath{\tau^+\tau^-}\xspace}
\newcommand{\wangle}{\ensuremath{\sin^{2}\theta_{\text{eff}}^\text{lept}(M^2_\mathrm{Z})}\xspace}
\newcommand{\ttbar}{\ensuremath{{t\overline{t}}}\xspace}
\newcommand{\stat}{\ensuremath{\,\text{(stat.)}}\xspace}
\newcommand{\syst}{\ensuremath{\,\text{(syst.)}}\xspace}

\newcommand{\HGG}{\ensuremath{\mathrm{H}\to\gamma\gamma}}
\newcommand{\gev}{\GeV}
\newcommand{\GAMJET}{\ensuremath{\gamma + \text{jet}}}
\newcommand{\PPTOJETS}{\ensuremath{\mathrm{pp}\to\text{jets}}}
\newcommand{\PPTOGG}{\ensuremath{\mathrm{pp}\to\gamma\gamma}}
\newcommand{\PPTOGAMJET}{\ensuremath{\mathrm{pp}\to\gamma +
\mathrm{jet}
}}
\newcommand{\MH}{\ensuremath{\mathrm{M_{\mathrm{H}}}}}
\newcommand{\RNINE}{\ensuremath{\mathrm{R}_\mathrm{9}}}
\newcommand{\DR}{\ensuremath{\Delta\mathrm{R}}}


\newcommand{\PT}{\ensuremath{p_{\mathrm{T}}}\xspace}
\newcommand{\pt}{\ensuremath{p_{\mathrm{T}}}\xspace}
\newcommand{\ET}{\ensuremath{E_{\mathrm{T}}}\xspace}
\newcommand{\HT}{\ensuremath{H_{\mathrm{T}}}\xspace}
\newcommand{\et}{\ensuremath{E_{\mathrm{T}}}\xspace}
\newcommand{\Em}{\ensuremath{E\!\!\!/}\xspace}
\newcommand{\Pm}{\ensuremath{p\!\!\!/}\xspace}
\newcommand{\PTm}{\ensuremath{{p\!\!\!/}_{\mathrm{T}}}\xspace}
\newcommand{\ETm}{\ensuremath{E_{\mathrm{T}}^{\text{miss}}}\xspace}
\newcommand{\MET}{\ensuremath{E_{\mathrm{T}}^{\text{miss}}}\xspace}
\newcommand{\ETmiss}{\ensuremath{E_{\mathrm{T}}^{\text{miss}}}\xspace}
\newcommand{\VEtmiss}{\ensuremath{{\vec E}_{\mathrm{T}}^{\text{miss}}}\xspace}

%

\newcommand{\ga}{\ensuremath{\gtrsim}}
\newcommand{\la}{\ensuremath{\lesssim}}
\newcommand{\swsq}{\ensuremath{\sin^2\theta_W}\xspace}
\newcommand{\cwsq}{\ensuremath{\cos^2\theta_W}\xspace}
\newcommand{\tanb}{\ensuremath{\tan\beta}\xspace}
\newcommand{\tanbsq}{\ensuremath{\tan^{2}\beta}\xspace}
\newcommand{\sidb}{\ensuremath{\sin 2\beta}\xspace}
\newcommand{\alpS}{\ensuremath{\alpha_S}\xspace}
\newcommand{\alpt}{\ensuremath{\tilde{\alpha}}\xspace}

\newcommand{\QL}{\ensuremath{Q_L}\xspace}
\newcommand{\sQ}{\ensuremath{\tilde{Q}}\xspace}
\newcommand{\sQL}{\ensuremath{\tilde{Q}_L}\xspace}
\newcommand{\ULC}{\ensuremath{U_L^C}\xspace}
\newcommand{\sUC}{\ensuremath{\tilde{U}^C}\xspace}
\newcommand{\sULC}{\ensuremath{\tilde{U}_L^C}\xspace}
\newcommand{\DLC}{\ensuremath{D_L^C}\xspace}
\newcommand{\sDC}{\ensuremath{\tilde{D}^C}\xspace}
\newcommand{\sDLC}{\ensuremath{\tilde{D}_L^C}\xspace}
\newcommand{\LL}{\ensuremath{L_L}\xspace}
\newcommand{\sL}{\ensuremath{\tilde{L}}\xspace}
\newcommand{\sLL}{\ensuremath{\tilde{L}_L}\xspace}
\newcommand{\ELC}{\ensuremath{E_L^C}\xspace}
\newcommand{\sEC}{\ensuremath{\tilde{E}^C}\xspace}
\newcommand{\sELC}{\ensuremath{\tilde{E}_L^C}\xspace}
\newcommand{\sEL}{\ensuremath{\tilde{E}_L}\xspace}
\newcommand{\sER}{\ensuremath{\tilde{E}_R}\xspace}
\newcommand{\sFer}{\ensuremath{\tilde{f}}\xspace}
\newcommand{\sQua}{\ensuremath{\tilde{q}}\xspace}
\newcommand{\sUp}{\ensuremath{\tilde{u}}\xspace}
\newcommand{\suL}{\ensuremath{\tilde{u}_L}\xspace}
\newcommand{\suR}{\ensuremath{\tilde{u}_R}\xspace}
\newcommand{\sDw}{\ensuremath{\tilde{d}}\xspace}
\newcommand{\sdL}{\ensuremath{\tilde{d}_L}\xspace}
\newcommand{\sdR}{\ensuremath{\tilde{d}_R}\xspace}
\newcommand{\sTop}{\ensuremath{\tilde{t}}\xspace}
\newcommand{\stL}{\ensuremath{\tilde{t}_L}\xspace}
\newcommand{\stR}{\ensuremath{\tilde{t}_R}\xspace}
\newcommand{\stone}{\ensuremath{\tilde{t}_1}\xspace}
\newcommand{\sttwo}{\ensuremath{\tilde{t}_2}\xspace}
\newcommand{\sBot}{\ensuremath{\tilde{b}}\xspace}
\newcommand{\sbL}{\ensuremath{\tilde{b}_L}\xspace}
\newcommand{\sbR}{\ensuremath{\tilde{b}_R}\xspace}
\newcommand{\sbone}{\ensuremath{\tilde{b}_1}\xspace}
\newcommand{\sbtwo}{\ensuremath{\tilde{b}_2}\xspace}
\newcommand{\sLep}{\ensuremath{\tilde{l}}\xspace}
\newcommand{\sLepC}{\ensuremath{\tilde{l}^C}\xspace}
\newcommand{\sEl}{\ensuremath{\tilde{e}}\xspace}
\newcommand{\sElC}{\ensuremath{\tilde{e}^C}\xspace}
\newcommand{\seL}{\ensuremath{\tilde{e}_L}\xspace}
\newcommand{\seR}{\ensuremath{\tilde{e}_R}\xspace}
\newcommand{\snL}{\ensuremath{\tilde{\nu}_L}\xspace}
\newcommand{\sMu}{\ensuremath{\tilde{\mu}}\xspace}
\newcommand{\sNu}{\ensuremath{\tilde{\nu}}\xspace}
\newcommand{\sTau}{\ensuremath{\tilde{\tau}}\xspace}
\newcommand{\Glu}{\ensuremath{g}\xspace}
\newcommand{\sGlu}{\ensuremath{\tilde{g}}\xspace}
\newcommand{\Wpm}{\ensuremath{W^{\pm}}\xspace}
\newcommand{\sWpm}{\ensuremath{\tilde{W}^{\pm}}\xspace}
\newcommand{\Wz}{\ensuremath{W^{0}}\xspace}
\newcommand{\sWz}{\ensuremath{\tilde{W}^{0}}\xspace}
\newcommand{\sWino}{\ensuremath{\tilde{W}}\xspace}
\newcommand{\Bz}{\ensuremath{B^{0}}\xspace}
\newcommand{\sBz}{\ensuremath{\tilde{B}^{0}}\xspace}
\newcommand{\sBino}{\ensuremath{\tilde{B}}\xspace}
\newcommand{\Zz}{\ensuremath{Z^{0}}\xspace}
\newcommand{\sZino}{\ensuremath{\tilde{Z}^{0}}\xspace}
\newcommand{\sGam}{\ensuremath{\tilde{\gamma}}\xspace}
\newcommand{\chiz}{\ensuremath{\tilde{\chi}^{0}}\xspace}
\newcommand{\chip}{\ensuremath{\tilde{\chi}^{+}}\xspace}
\newcommand{\chim}{\ensuremath{\tilde{\chi}^{-}}\xspace}
\newcommand{\chipm}{\ensuremath{\tilde{\chi}^{\pm}}\xspace}
\newcommand{\Hone}{\ensuremath{H_{d}}\xspace}
\newcommand{\sHone}{\ensuremath{\tilde{H}_{d}}\xspace}
\newcommand{\Htwo}{\ensuremath{H_{u}}\xspace}
\newcommand{\sHtwo}{\ensuremath{\tilde{H}_{u}}\xspace}
\newcommand{\sHig}{\ensuremath{\tilde{H}}\xspace}
\newcommand{\sHa}{\ensuremath{\tilde{H}_{a}}\xspace}
\newcommand{\sHb}{\ensuremath{\tilde{H}_{b}}\xspace}
\newcommand{\sHpm}{\ensuremath{\tilde{H}^{\pm}}\xspace}
\newcommand{\hz}{\ensuremath{h^{0}}\xspace}
\newcommand{\Hz}{\ensuremath{H^{0}}\xspace}
\newcommand{\Az}{\ensuremath{A^{0}}\xspace}
\newcommand{\Hpm}{\ensuremath{H^{\pm}}\xspace}
\newcommand{\sGra}{\ensuremath{\tilde{G}}\xspace}
\newcommand{\mtil}{\ensuremath{\tilde{m}}\xspace}
\newcommand{\rpv}{\ensuremath{\rlap{\kern.2em/}R}\xspace}
\newcommand{\LLE}{\ensuremath{LL\bar{E}}\xspace}
\newcommand{\LQD}{\ensuremath{LQ\bar{D}}\xspace}
\newcommand{\UDD}{\ensuremath{\overline{UDD}}\xspace}
\newcommand{\Lam}{\ensuremath{\lambda}\xspace}
\newcommand{\Lamp}{\ensuremath{\lambda'}\xspace}
\newcommand{\Lampp}{\ensuremath{\lambda''}\xspace}
\newcommand{\spinbd}[2]{\ensuremath{\bar{#1}_{\dot{#2}}}\xspace}

\newcommand{\MD}{\ensuremath{{M_\mathrm{D}}}\xspace}
\newcommand{\Mpl}{\ensuremath{{M_\mathrm{Pl}}}\xspace}
\newcommand{\Rinv} {\ensuremath{{R}^{-1}}\xspace}

%
%
\hyphenation{en-viron-men-tal}

%
%
\newcommand {\roots}    {\ensuremath{\sqrt{s}}}
\newcommand {\dndy}     {\ensuremath{dN/dy}}
\newcommand {\dnchdy}   {\ensuremath{dN_{\mathrm{ch}}/dy}}
\newcommand {\dndeta}   {\ensuremath{dN/d\eta}}
\newcommand {\dnchdeta} {\ensuremath{dN_{\mathrm{ch}}/d\eta}}
\newcommand {\dndpt}    {\ensuremath{dN/d\pt}}
\newcommand {\dnchdpt}  {\ensuremath{dN_{\mathrm{ch}}/d\pt}}
\newcommand {\deta}     {\ensuremath{\Delta\eta}}
\newcommand {\dphi}     {\ensuremath{\Delta\phi}}

\newcommand {\pp}    {\mbox{pp}}
\newcommand {\ppbar} {\mbox{p\={p}}}
\newcommand {\pbarp} {\mbox{p\={p}}}
\newcommand {\PbPb}  {\mbox{PbPb }}

\newcommand{\m}{\ensuremath{\,\text{m}}\xspace}

\newcommand {\naive}    {na\"{\i}ve}
\providecommand{\GEANT} {{Geant}\xspace}
\providecommand{\PHOJET} {\textsc{phojet}\xspace}

\def\d{\mathrm{d}}

\providecommand{\PKzS}{\ensuremath{\mathrm{K^0_S}}}
\providecommand{\Pp}{\ensuremath{\mathrm{p}}}
\providecommand{\Pap}{\ensuremath{\mathrm{\overline{p}}}}
\providecommand{\PgL}{\ensuremath{\Lambda}}
\providecommand{\PagL}{\ensuremath{\overline{\Lambda}}}
\providecommand{\PgS}{\ensuremath{\Sigma}}
\providecommand{\PgSm}{\ensuremath{\Sigma^-}}
\providecommand{\PgSp}{\ensuremath{\Sigma^+}}
\providecommand{\PagSm}{\ensuremath{\overline{\Sigma}^-}}
\providecommand{\PagSp}{\ensuremath{\overline{\Sigma}^+}}
\renewcommand{\GEANTfour} {{Geant4}\xspace}

\cmsNoteHeader{QCD-09-010} 
\title{Transverse-momentum and pseudorapidity distributions of charged
hadrons in \pp\ collisions at $\sqrt{s}$~=~0.9 and 2.36~TeV}

\address[cern]{CERN}
\author[cern]{CMS Collaboration}

\date{\today}

\abstract{ 
Measurements of inclusive charged-hadron transverse-momentum and pseudorapidity 
distributions are presented for proton-proton collisions at $\sqrt{s}=$~0.9 and 
2.36~TeV. The data were collected with the CMS detector 
during the LHC commissioning in December 2009. 
For non-single-diffractive interactions, the average charged-hadron transverse momentum is 
measured to be 
$0.46 \pm 0.01$~(stat.)~$\pm$~0.01~(syst.)~\GeVc\ at 0.9~TeV and 
$0.50 \pm 0.01$~(stat.)~$\pm$~0.01~(syst.)~\GeVc\ at 2.36~TeV, 
for pseudorapidities between $-2.4$ and $+2.4$. 
At these energies, the measured pseudorapidity 
densities in the central region, $\dnchdeta |_{|\eta| < 0.5}$, are 
$3.48 \pm 0.02$~(stat.)~$\pm$~0.13~(syst.) and
$4.47 \pm 0.04$~(stat.)~$\pm$~0.16~(syst.), respectively.
The results at 0.9~TeV are in agreement with previous measurements 
and confirm the expectation of near equal hadron production in \ppbar\ and 
\pp\ collisions. The results at 2.36~TeV represent the highest-energy 
measurements at a particle collider to date.
}

\hypersetup{%
pdfauthor={
Ivan Cali, Tae Jeong Kim, Yongsun Kim, Krisztian Krajczar, Yen-Jie Lee, 
Wei Li, Constantin Loizides, Frank Ma, Christof Roland, Gunther Roland, 
Romain Rougny, Ferenc Sikler, Hella Snoek, Gabor Veres, Edward Wenger, 
Yetkin Yilmaz, Andre Yoon. Editor: Vivek Sharma},%
pdftitle={Transverse momentum and pseudorapidity distributions of charged
hadrons in \pp\ collisions at sqrt s~=~0.9 and 2.36~TeV},%
pdfsubject={CMS},%
pdfkeywords={CMS, multiplicity, transverse momentum, soft QCD, NSD, physics}}

\maketitle 


\section{Introduction}
\label{sect:intro}

Measurements of transverse-momentum ($\pt$) and pseudorapidity ($\eta$)
distributions are reported for charged hadrons produced in 
proton-proton (\pp) collisions 
at centre-of-mass energies ($\roots$) of 0.9 and 2.36~TeV at the CERN Large 
Hadron Collider (LHC) \cite{Evans:2008zzb}. The data were recorded
with the Compact Muon Solenoid (CMS) experiment in December 2009
during two 2-hour periods of the LHC commissioning, demonstrating
the readiness of CMS in the early phase of LHC operations.
The results at $\roots = 2.36$~TeV represent the highest-energy measurements 
at a particle collider to date.

The majority of \pp\ collisions are soft, i.e., without any hard 
scattering of the partonic constituents of the proton. In contrast to the 
higher-\pt regime, well described by perturbative QCD, particle production 
in soft collisions is generally modelled phenomenologically to describe 
the different \pp\ scattering processes: elastic scattering, 
single-diffractive and double-diffractive dissociation, and inelastic 
non-diffractive scattering~\cite{Kittel:2005fu}.

The measurements presented in this paper are the inclusive primary 
charged-hadron multiplicity densities (\dnchdpt\ and \dnchdeta) 
in the pseudorapidity range $|\eta| < 2.4$, where $\pt$ is the momentum of 
the particle transverse to the beam axis, 
and where $N_{\rm ch}$ is the number of charged hadrons in any given 
$\eta$ or $\pt$ interval. The pseudorapidity $\eta$ is defined as 
$-\ln[\tan(\theta/2)]$, where $\theta$ is the polar angle of the particle 
with respect to the anti-clockwise beam direction.

Primary charged hadrons are defined as all charged hadrons 
produced in the interactions, including the products of strong and 
electromagnetic decays, but excluding products of weak decays 
and hadrons originating from secondary interactions. 
In this paper, the multiplicity densities are measured for inelastic 
non-single-diffractive (NSD) interactions to minimize the model dependence 
of the necessary corrections for the event selection, and to enable a 
comparison with earlier experiments. The event selection was therefore 
designed to retain a large fraction of inelastic double-diffractive (DD) 
and non-diffractive (ND) events, while rejecting all elastic and most 
single-diffractive dissociation (SD) events.

Measurements of \dnchdpt\ and \dnchdeta\ distributions and their $\roots$ 
dependence are important for understanding the mechanisms of hadron 
production and the relative roles of soft and hard scattering 
contributions in the LHC energy regime.
Furthermore, the measurements at the highest collision energy of 2.36~TeV
are a first step towards understanding inclusive particle production at a 
new energy frontier. These measurements will be particularly relevant for 
the LHC as, when it is operated at design luminosity, rare signal events 
will be embedded in a background of more than 20 near-simultaneous 
minimum-bias collisions. These results will also serve as a reference in 
the measurement of nuclear-medium effects in \PbPb\,collisions at the LHC.
The differences in these distributions between \pp\ and \pbarp\ 
collisions are expected to be smaller than the attainable precision 
of these measurements~\cite{Rushbrooke:1977ni}.
The results reported here at $\roots = 0.9$~TeV can therefore be directly compared to 
those previously obtained in \pbarp\ collisions.

This paper is organized as follows. In Section~\ref{sect:detector}, the 
elements of the CMS detector relevant to this analysis are outlined.
In Sections~\ref{sect:evtSel} and \ref{sect:reco}, the event selection 
and reconstruction algorithms are described. Results on \dnchdpt\ and 
\dnchdeta\ are presented in Section~\ref{sect:results} and compared 
with previous \ppbar\ and \pp\ measurements in Section~\ref{sect:discuss}.

\section{The CMS detector}
\label{sect:detector}

A detailed description of the CMS experiment can be found in Ref.~\cite{JINST}.
The central feature of the CMS apparatus is a superconducting solenoid
of 6~m internal diameter, providing a uniform magnetic field of 3.8 T.
Immersed in the magnetic field are the pixel tracker, the silicon-strip 
tracker (SST), the lead-tungstate crystal electromagnetic calorimeter 
(ECAL) and the brass/scintillator hadron calorimeter (HCAL). 
In addition to barrel and end-cap detectors for ECAL and HCAL, 
the steel/quartz-fibre forward calorimeter (HF) covers the region of 
$|\eta|$ between 2.9 and 5.2. The HF tower segmentation in $\eta$ and azimuthal angle $\phi$
(expressed in radians) is 0.175$\times$0.175, except for $|\eta|$ above 4.7 where the segmentation 
is 0.175$\times$0.35. Muons are measured in gas-ionization 
detectors embedded in the steel return yoke. 

The tracker consists of 1440 silicon-pixel and 15\,148 silicon-strip
detector modules and measures charged particle trajectories within the nominal
pseudorapidity range $|\eta|< 2.5$. The pixel tracker consists of three
53.3 cm long barrel layers and two end-cap disks on each side of the barrel
section. The innermost barrel layer has a radius of 4.4 cm, while for the
second and third layers the radii are 7.3 cm and 10.2 cm, respectively.
The tracker is designed to provide an impact-parameter resolution of about
100~$\mu$m and a transverse-momentum resolution of about 0.7\,\% for
1~\GeVc charged particles at normal incidence ($\eta$=0)~\cite{CMSTDR1}.

During the data-taking period addressed by this analysis, 98.4\% of 
the pixel and 97.2\% of the SST channels were operational.
The fraction of noisy pixel channels was less than $10^{-5}$.
The signal-to-noise ratio in the SST depends on the sensor thickness and was
measured to be between 28 and 36, consistent with the 
design expectations and cosmic-ray measurements~\cite{Collaboration:2009dv,CMSTDR1}.
The tracker was aligned as described in Ref.~\cite{TrackerAlign} using 
cosmic ray data prior to the LHC commissioning. The 
precision achieved for the positions of the detector modules with respect 
to particle trajectories is 3-4~$\mu$m in the barrel for the coordinate 
in the bending plane.


Two elements of the CMS detector monitoring system, the Beam Scintillator Counters 
(BSCs)~\cite{JINST, Bell} and the Beam Pick-up Timing for the eXperiments 
(BPTX) devices~\cite{JINST, Aumeyr}, were used to trigger the detector readout.
The two BSCs are located at a distance of 
$\pm 10.86$~\m from the nominal interaction point (IP) and are sensitive 
in the $|\eta|$ range from 3.23 to 4.65. 
Each BSC is a set of $16$ scintillator tiles. The 
BSC elements have a time resolution of 3~ns and an average 
minimum-ionizing-particle detection efficiency of 96.3\%, and are designed 
to provide hit and coincidence rates. 
The two BPTX devices,
located around the beam pipe at a distance of $\pm 175$~m from the IP on 
either side, are designed to provide precise information on the bunch 
structure and timing of the incoming beam, with better than 0.2 ns time 
resolution.

The CMS experiment uses a right-handed coordinate system, with the origin at 
the nominal interaction point, the $x$ axis pointing to the centre of the 
LHC, the $y$ axis pointing up (perpendicular to the LHC plane) and the 
$z$ axis along the anticlockwise-beam direction. The azimuthal angle, 
$\phi$, is measured in the ($x$,$y$) plane, where $\phi=0$ is the $+x$ 
and $\phi=\pi/2$ is the $+y$ direction. 

The detailed Monte Carlo (MC) simulation of the CMS detector response is based 
on \GEANTfour\ ~\cite{GEANT4}. The position and width of the beam spot in 
the simulation were adjusted to that determined from the data. Simulated 
events were processed and reconstructed in the same manner as collision 
data.

\section{Event selection}
\label{sect:evtSel} 

This analysis uses two LHC collision data sets collected with 
\pp\ interaction rates of about 11 and 3 Hz at $\roots = 0.9$ and 2.36 TeV, 
respectively. At these rates, the probability for more than one inelastic 
collision to occur in the same proton bunch crossing is less than $2\times 10^{-4}$
at both collision energies.

Events were selected by a trigger signal in any of the BSC scintillators, 
coincident with a signal from either of the two BPTX detectors indicating 
the presence of at least one proton bunch crossing the IP. 
From these samples, collision events were selected offline by requiring BPTX signals from 
both beams passing the IP and at least one reconstructed charged particle 
trajectory in the pixel detector originating from within 0.2 cm of the 
beam position in the transverse direction (Section \ref{sect:reco_vertex}). 
The total number of collision events and the numbers of collision events 
passing each requirement are listed in Table \ref{tab:eventnum}.
\begin{table}[tbh]
  \caption{Numbers of events per data sample used in this analysis. The 
offline event selection criteria are applied in sequence, i.e., each line 
includes the selection of the lines above. }
  \label{tab:eventnum}
  \begin{center}
    \begin{tabular}{lcc}
\hline
\hline
Centre-of-mass Energy     &  0.9 TeV     & 2.36 TeV \\
\hline
Selection       & \multicolumn{2}{c}{Number of Events} \\
\hline
BPTX Coincidence + one BSC Signal & 72\,637 & 18\,074 \\
One Pixel Track    & 51\,308 & 13\,029 \\
HF Coincidence     & 40\,781 & 10\,948 \\
Beam Halo Rejection & 40\,741 & 10\,939  \\
Beam Background Rejection & 40\,647 & 10\,905 \\
Valid Event Vertex & 40\,320 & 10\,837 \\
\hline
\hline
    \end{tabular}
  \end{center}
\end{table}

To select NSD events, a coincidence of at least one 
HF calorimeter tower with more than 3 GeV total energy 
on each of the positive and negative sides of the HF was required.
%
%
Events containing beam-halo muons crossing the detector were identified
by requiring the time difference between any two hits from the BSC stations on 
opposite sides of the IP to be within $73\pm 20$~ns. Such events were removed 
from the data sample.
%
%
Beam-induced background events producing an anomalously large number of pixel hits 
 were excluded by rejecting events with pixel clusters (Section~\ref{sect:reco_clusters})
inconsistent with a \pp\ collision vertex.
This rejection algorithm was only applied for events 
with more than 150 pixel clusters, providing a clean separation between 
collision events and beam background events. 
Finally, events were required to contain at least one reconstructed primary 
vertex, as described in Section~\ref{sect:reco}. 

To study beam-induced background, the event selection criteria 
were also applied to a data sample obtained by selecting events
with only a single unpaired bunch crossing the IP. The contamination 
of background events in the colliding-bunch data sample was estimated 
by taking into account the total unpaired and paired bunch intensities 
and was found to be negligible ($<$0.1\%). 
The total number of cosmic-ray muons in the selected data sample was 
estimated to be less than one event, and was also neglected.

The event selection criteria are expected to have high efficiency for the NSD part of 
the \pp\ cross section, while rejecting a large fraction of the SD component
of \pp\ interactions. The efficiency of the event selection for the different 
processes and centre-of-mass energies was determined using simulated 
events obtained from the \PYTHIA~\cite{Sjostrand:2006za} (version 6.420, tune 
D6T,~\cite{Bartalini:2009xx}) and \PHOJET~\cite{Bopp:1998rc, Engel:1995sb} 
(version 1.12-35) event generators processed with a MC simulation of the CMS 
detector response (hereafter simply called \PYTHIA~and \PHOJET). 
In the case of \PHOJET, the discussion and numerical values concerning 
the DD process given in this paper contain both the DD and the 
Double-Pomeron-Exchange (DPE) processes.
The relative event fractions of SD, DD and ND 
processes and event selection efficiencies at $\roots =$ 0.9 and 2.36 TeV 
are listed in Table \ref{tab:difffrac} for these two samples. 

The measurements were corrected for the selection efficiency of NSD processes
and for the fraction of SD events contained in the data sample after the 
event selection. Based on the \PYTHIA (\PHOJET) event generator, the 
fractions of SD events contained in the selected data samples were 
estimated to be 5.2\% (4.9\%) at 0.9 TeV and 6.3\% (5.0\%) at 2.36 TeV.

\begin{table}[tbh]
  \caption{Expected fractions of SD, DD, ND and NSD processes 
(``Frac.'')~obtained from the \PYTHIA~and \PHOJET~event generators before any selection 
and the corresponding selection efficiencies (``Sel. Eff.'') determined 
from the MC simulation. 
}
  \label{tab:difffrac}
  \begin{center}
    \begin{tabular}{lcc|cc||cc|cc}
\hline
\hline
  	& \multicolumn{4}{c||}{\PYTHIA} & \multicolumn{4}{c}{\PHOJET} \\
\hline
Energy     & \multicolumn{2}{c|}{ 0.9 TeV} & \multicolumn{2}{c||}{2.36 TeV} & \multicolumn{2}{c|}{ 0.9 TeV} & \multicolumn{2}{c}{2.36 TeV} \\
		& Frac. & Sel. Eff. & Frac. & Sel. Eff. & Frac. & Sel. Eff. & Frac. & Sel. Eff.  \\
\hline
 SD             & 22.5\%    & 16.1\%    & 21.0\%    & 21.8\% & 18.9\% & 20.1\% & 16.2\% & 25.1\% \\
 DD             & 12.3\%    & 35.0\%    & 12.8\%    & 33.8\% &  8.4\% & 53.8\% &  7.3\% & 50.0\% \\
 ND             & 65.2\%    & 95.2\%    & 66.2\%    & 96.4\% & 72.7\% & 94.7\% & 76.5\% & 96.5\% \\
 NSD            & 77.5\%    & 85.6\%    & 79.0\%    & 86.2\% & 81.1\% & 90.5\% & 83.8\% & 92.4\% \\
\hline
\hline
    \end{tabular}
  \end{center}
\end{table}

\begin{figure}
 \begin{flushleft}
  \subfigure{\label{fig:trigeff_a}
  \includegraphics[width=0.45\textwidth]{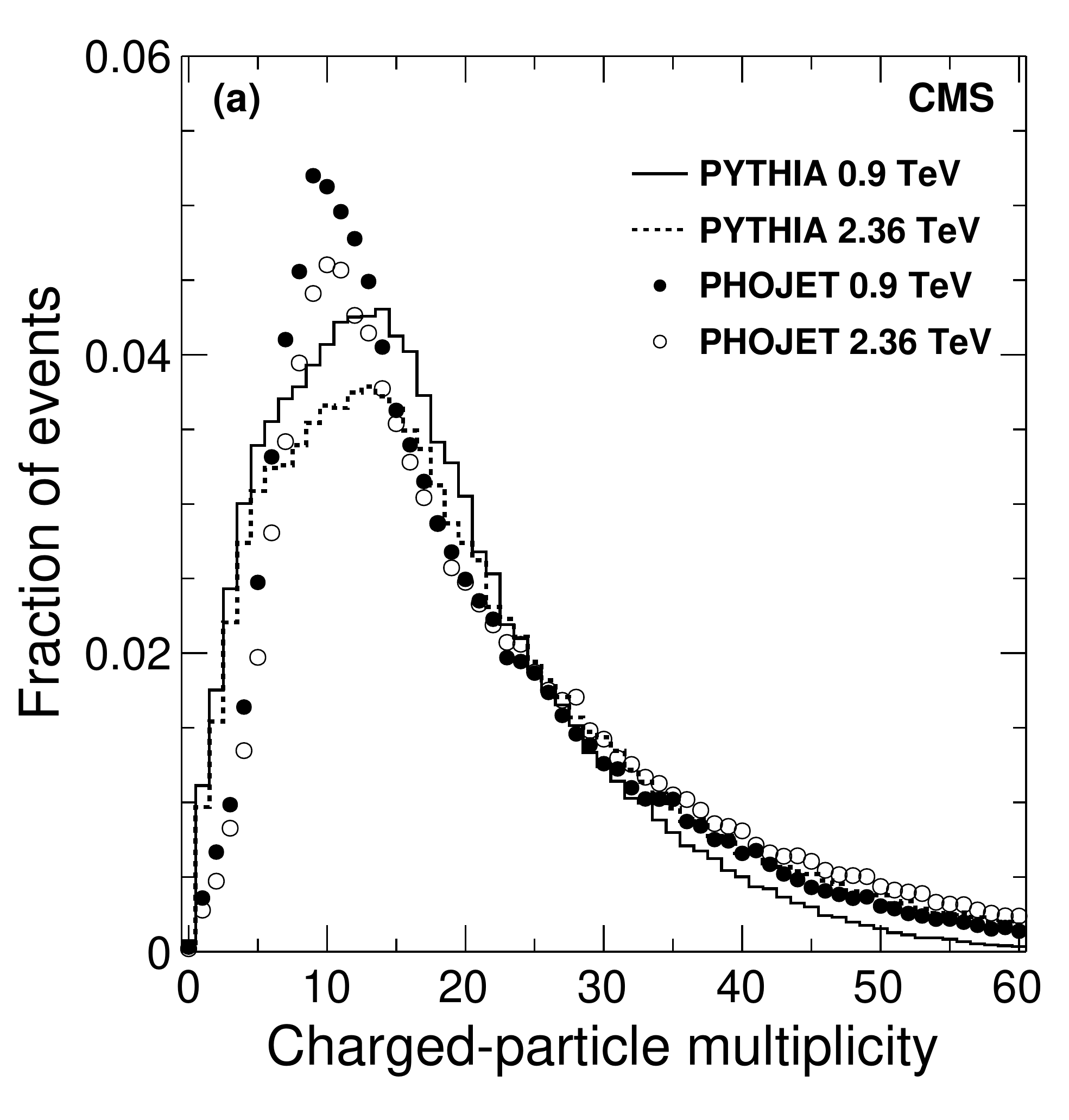}}
  \hspace{1cm}
  \subfigure{\label{fig:diffrac_b}
  \includegraphics[width=0.45\textwidth]{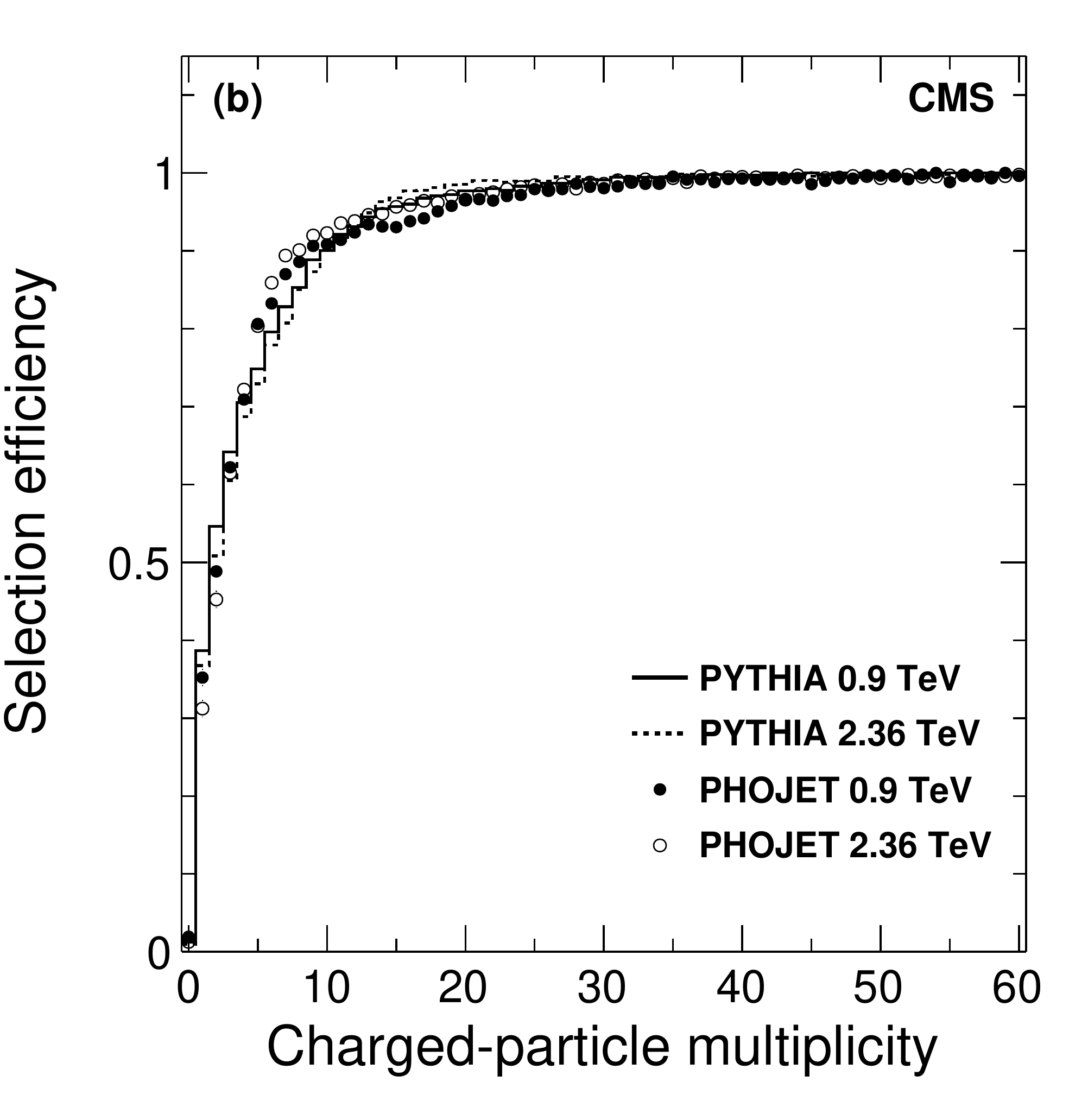}}
 \end{flushleft}
 \caption{
(a) Generated multiplicity distributions of primary charged hadrons in 
the range $|\eta| < 2.5$ for $\roots =$ 0.9 TeV (solid dots and histogram) and 2.36~TeV (open circles and dashed histogram) after the 
event selection is applied to the reconstructed events, using inelastic 
events from the \PYTHIA (histograms) and \PHOJET (symbols) event generators. 
(b) The event selection efficiency expected for NSD events 
from the \PYTHIA (histograms) and \PHOJET (symbols) event generators
as a function of generated charged hadron multiplicity in the region 
$|\eta|<2.5$.}
 \label{fig:trigeff}
\end{figure}

The generated charged-hadron multiplicity distribution is shown in 
Fig.~\ref{fig:trigeff}a in the range $|\eta|<$~2.5 for all inelastic 
events after event selection. The event selection efficiency for NSD 
events is shown in Fig.~\ref{fig:trigeff}b as a function of generated 
charged-hadron multiplicity in the region $|\eta|<$~2.5.
The correction for the event selection efficiency was applied as a 
function of number of reconstructed charged particles per event, as 
illustrated at generator level in Figs.~\ref{fig:trigeff}a 
and~\ref{fig:trigeff}b.

The sum of the corrections to the \dnchdeta\ measurements due to the NSD event
selection efficiency and the SD event contamination typically amounts to 8\%.
The corrections applied in the analysis are based on \PYTHIA using the default 
SD and DD process fractions as listed in Table~\ref{tab:difffrac}. 

The \PYTHIA predictions for the SD and DD fractions differ from those 
of \PHOJET, and are not fully consistent with existing measurements, as explained 
in Section~\ref{sect:results}. These differences propagate to a systematic uncertainty
of 2\% in the \dnchdeta\ measurement.
To estimate the additional systematic uncertainty on the event 
selection efficiency correction resulting from the possible inaccuracies in the detector simulation, 
the analysis was repeated after replacing the HF event-selection criterion with a two-sided 
hit coincidence of signals in the BSC detectors. Based on this comparison, 
an additional 1\% systematic uncertainty was assigned to the \dnchdeta\ measurements.


\section{Reconstruction algorithms}
\label{sect:reco}
The analysis presented in this paper measures the \dnchdeta\ and
\dnchdpt\ distributions of primary charged hadrons. 
The \dnchdeta\ distributions were
obtained using three methods based on counting of (i) 
reconstructed clusters in the pixel barrel detector; (ii) pixel 
tracklets composed of pairs of clusters in different pixel barrel layers; 
and (iii) tracks reconstructed in the full tracker volume, 
combining the pixel and strip hits. The cluster counting method provides 
an independent measurement for each pixel barrel layer, and the 
tracklet method for each pair of layers. The third method also allows a 
measurement of the \dnchdpt\ distribution. All three methods rely on the 
reconstruction of the primary vertex (PV) described in 
Section~\ref{sect:reco_vertex}.

The pixel-cluster-counting method has
the advantage of having the largest $p_T$ acceptance down to small transverse 
momentum (30~\MeVc ), is insensitive to geometrical misalignment of the 
detector and does not require detailed knowledge of the primary 
vertex position resolution. A potential disadvantage is the sensitivity to 
backgrounds from collisions with residual gas in the beam pipe (beam-gas 
collisions), from secondary particles produced in the detector material
and from low-\pt particles curling in the axial 
magnetic field (loopers). The pixel-tracklet method is capable of 
measuring and correcting for the combinatorial background and has a \pt 
threshold of 50 \MeVc. The third method uses the tracker 
(pixel and SST) to build tracks. It requires at least two pixel hits in 
different layers, has the largest \pt threshold ($\approx$100 
\MeVc) and algorithmic complexity, but is the most robust against background 
hits produced by particles not originating from the collision.
The charged-particle multiplicity was corrected in all three methods for 
the small contamination ($<1\%$) of primary charged leptons.
The measured \dnchdeta\ values were evaluated by extrapolating or 
correcting to $\pt =0$ for all the three analysis methods.

The three reconstruction methods are described in 
Sections~\ref{sect:reco_clusters}- \ref{sect:reco_tracks}.

\subsection{Primary vertex reconstruction}
\label{sect:reco_vertex} 

The $x,y$ and $z$ positions of the luminous region where protons of both 
beams interact, hereafter referred to as beam spot, are obtained 
for each data set from three-dimensional vertex fits
based on tracks reconstructed with $\pt>0.9$~GeV/c, using the full
event sample. The RMS of the beam spot in the transverse
directions was found to be less than 0.05~cm. 
The beam spot position and dimensions were found to be stable within a 
given data set.

To reconstruct the $z$ coordinate of the PV for each event, tracks 
consisting of triplets of pixel hits were formed. 
The minimum transverse momentum of these tracks is $\approx 75$
\MeVc. The tracks were required to originate from the vicinity of the
beam spot with a transverse impact parameter ($d_{\rm T}$) smaller than
0.2\cm. Of these tracks, only those with $d_{\rm T} < 4\,\sigma_{\rm T}$, where 
$\sigma_{\rm T}$ is the quadratic sum of the uncertainty in $d_{\rm T}$
and the RMS of the beam spot in the transverse direction, were used in the 
vertex reconstruction. 


The vertex-reconstruction algorithm uses the $z$ coordinate of the
tracks at the point of closest approach to the beam axis and the
corresponding estimated measurement uncertainty ($\sigma_z$). It
performs an agglomerative clustering by adding tracks to form groups.
These groups (denoted the $\it{i^{\rm th}}$ and $\it{j^{\rm th}}$ group)
are then merged based on their normalized distance, $d_{ij}^2 = (z_i -
z_j)^2/(\sigma_i^2 + \sigma_j^2)$ where $\sigma_i$ and $\sigma_j$ are
the uncertainties of the $z_i$ and $z_j$ positions, with a fast
nearest-neighbour search algorithm \cite{Sikler:2009nx}. The $z$
position and its uncertainty $\sigma_z$ for the newly joined group are
calculated using a weighted average. The clustering process stops when
the smallest normalized distance between the remaining groups gets
larger than 8.0, where the stopping condition was optimized using simulated
events. Only vertices formed from at least two tracks were considered
except when only one track was reconstructed in the event. In the latter 
case the PV position was defined as the point of the closest approach
of the track to the beam axis. The fraction of single-track vertices in
the selected data sample is 1.7\% (1.3\%) at 0.9 TeV (2.36 TeV). The overall PV
reconstruction efficiency, evaluated from the data after all other event
selection cuts, is in excess of 99\% and the fraction of events with more
than one primary vertex candidate is 5.0\% (7.4\%) at 0.9~TeV (2.36~TeV). 
In the rare case of multiple PV candidates, the vertex composed of the largest set 
of tracks was chosen. 

The reconstructed primary vertex resolution in the
$z$ direction is a function of the associated track multiplicity
($N$) and was found to be parameterized adequately as $0.087$~cm~$/{N}^{0.6}$ using simulated events.

The distribution of the reconstructed $z$ position of the PV is shown 
in Fig.~\ref{background}a. Overlaid is the PV distribution in simulated
events, the position and RMS of which were adjusted to reproduce the beam 
spot measured in data.

\subsection{Pixel cluster counting method}
\label{sect:reco_clusters}

\begin{figure}[tp]
\centering
\subfigure{
	\label{vz_npix_1}
	\includegraphics[width=0.45\textwidth]{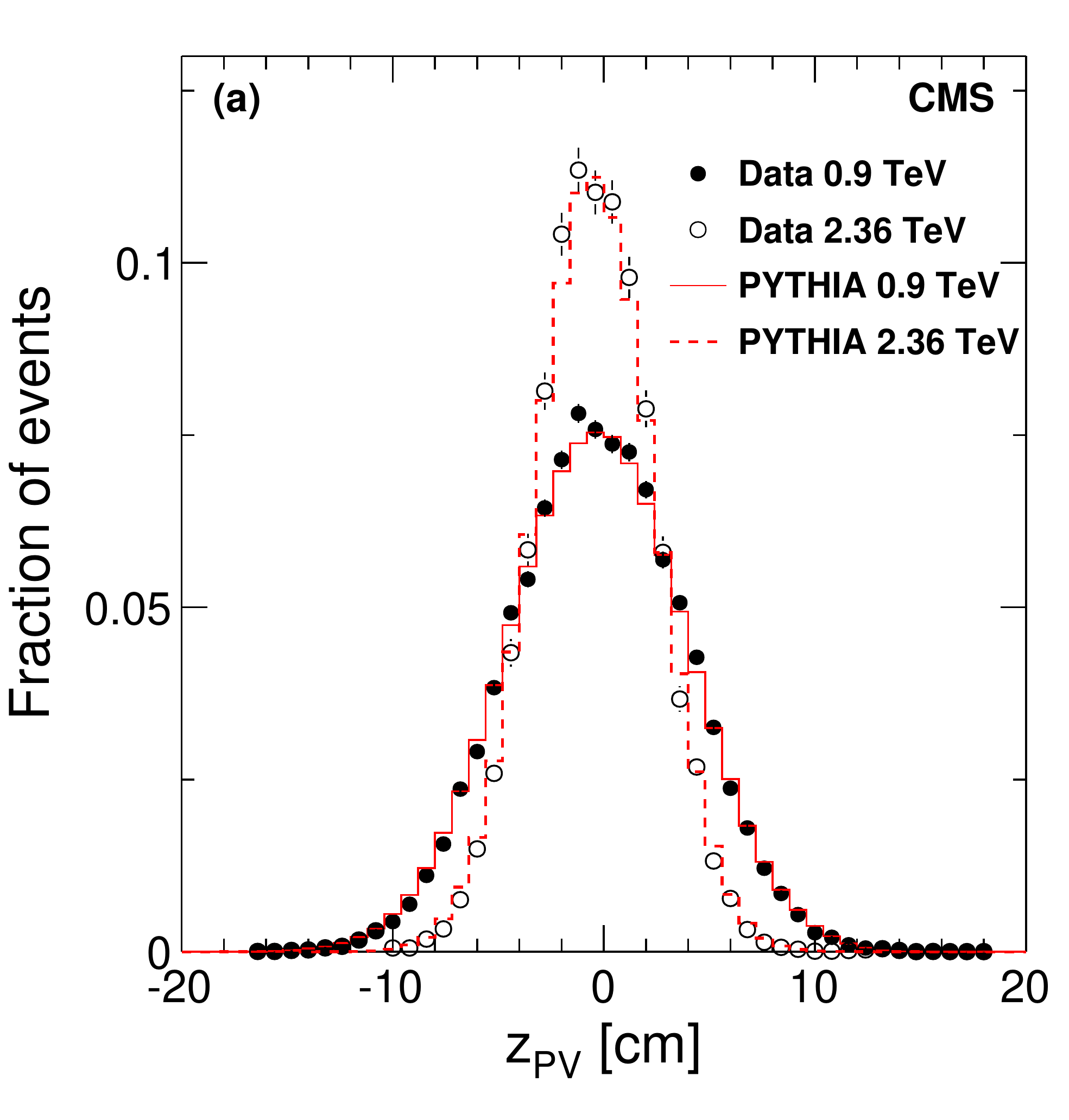}
}
\hspace{1cm}
\subfigure{
	\label{landau}
	\includegraphics[width=0.45\textwidth]{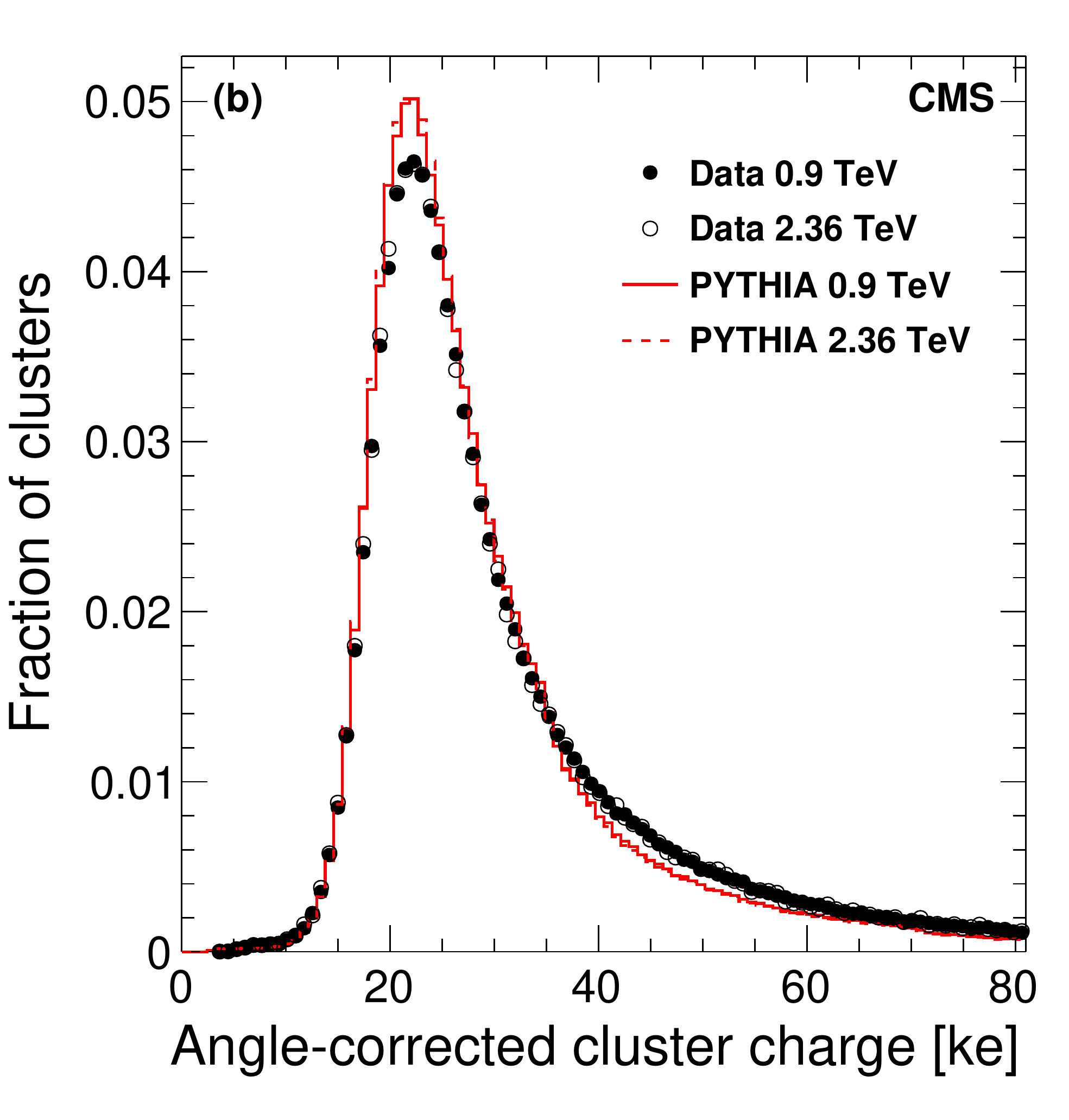}
}
\caption{
    \label{background}
	\subref{vz_npix_1} The distribution of the reconstructed $z$ 
	position of the primary vertex in the data (dots), compared to that from the \PYTHIA simulation (histogram).
	\subref{landau} Distribution of the cluster charge multiplied by 
	$|\sin\theta|$ in the data (dots) and the simulation (histogram),
	for the clusters selected for analysis. 
	}
\end{figure}


The pseudorapidity distribution of primary charged hadrons produced in a 
\pp\ collision can be measured by counting the number of clusters they 
create when traversing each of the three pixel barrel layers and applying 
appropriate corrections, as described in this section.

The energy deposited by charged particles traversing a pixel detector layer 
is spread over multiple pixels. Adjacent pixels with a charge measurement 
above a readout threshold of typically 2740 electrons are combined into pixel 
clusters to integrate the total charge 
deposit~\cite{Collaboration:2009dv}. The cluster size and charge
depend on the incident angle of the particle with respect to the active 
detector surface. The cluster length along the $z$ axis ranges from 1-2 pixels
at normal incident angle up to 14-16 pixels at shallow crossing angles.
Figure~\ref{background}b shows the measured distribution of cluster charge 
multiplied by $|\sin\theta|$ (or $1/\!\cosh\eta$) after the cluster selection
discussed below, compared to 
the simulation. Here, $\theta$ is the polar angle of the straight 
line connecting the PV to the cluster. 

The peak position is consistent with the expected charge of 
22\,ke, while the width of the distribution is slightly larger in the data 
than in the simulation due to gain calibration non-uniformities.

\begin{figure}[tp]
\centering
\subfigure{
	\label{background_1}
	\includegraphics[width=0.45\textwidth]{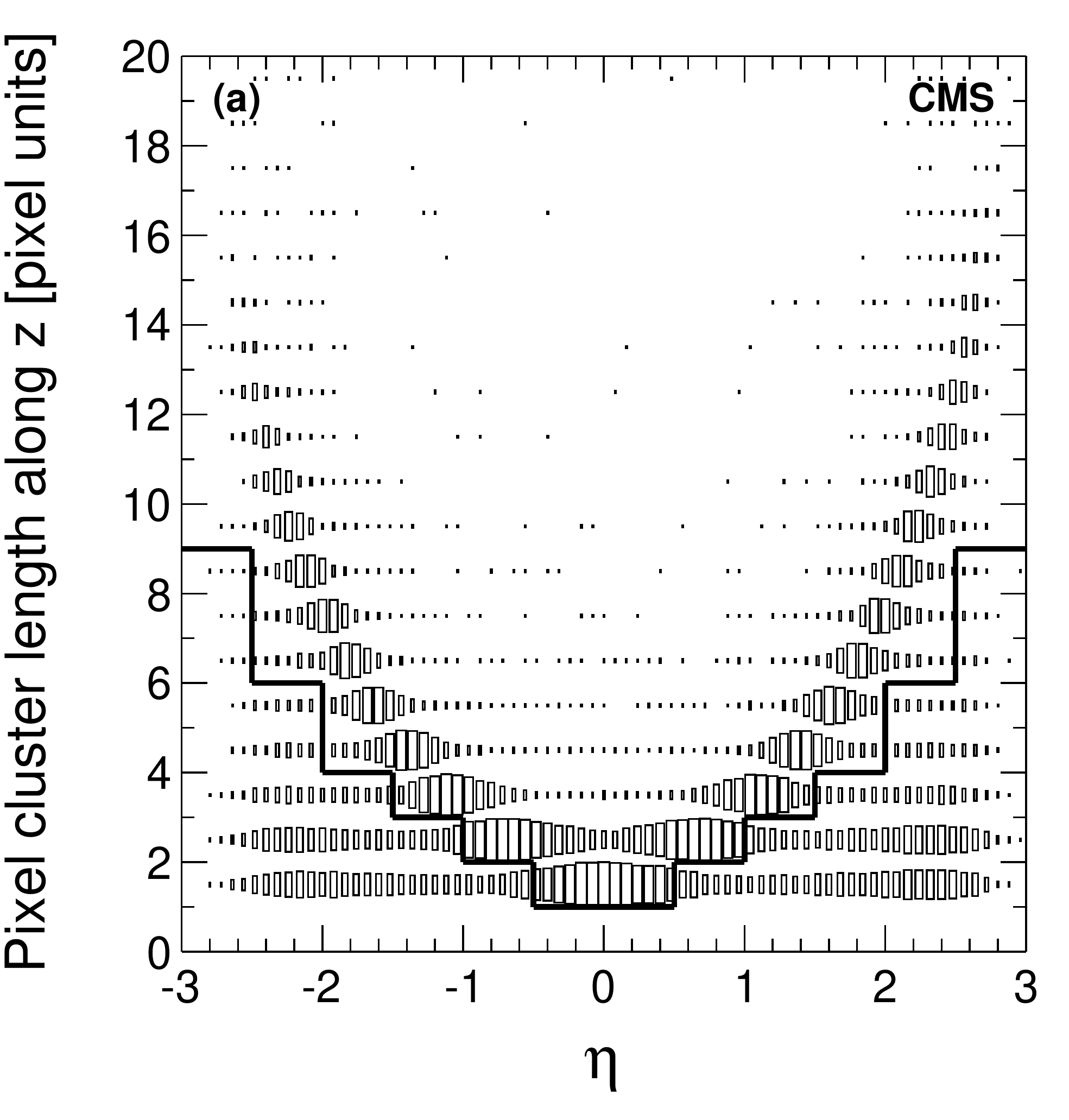}
}
\hspace{1cm}
\subfigure{
	\label{background_2}
        \includegraphics[width=0.45\textwidth]{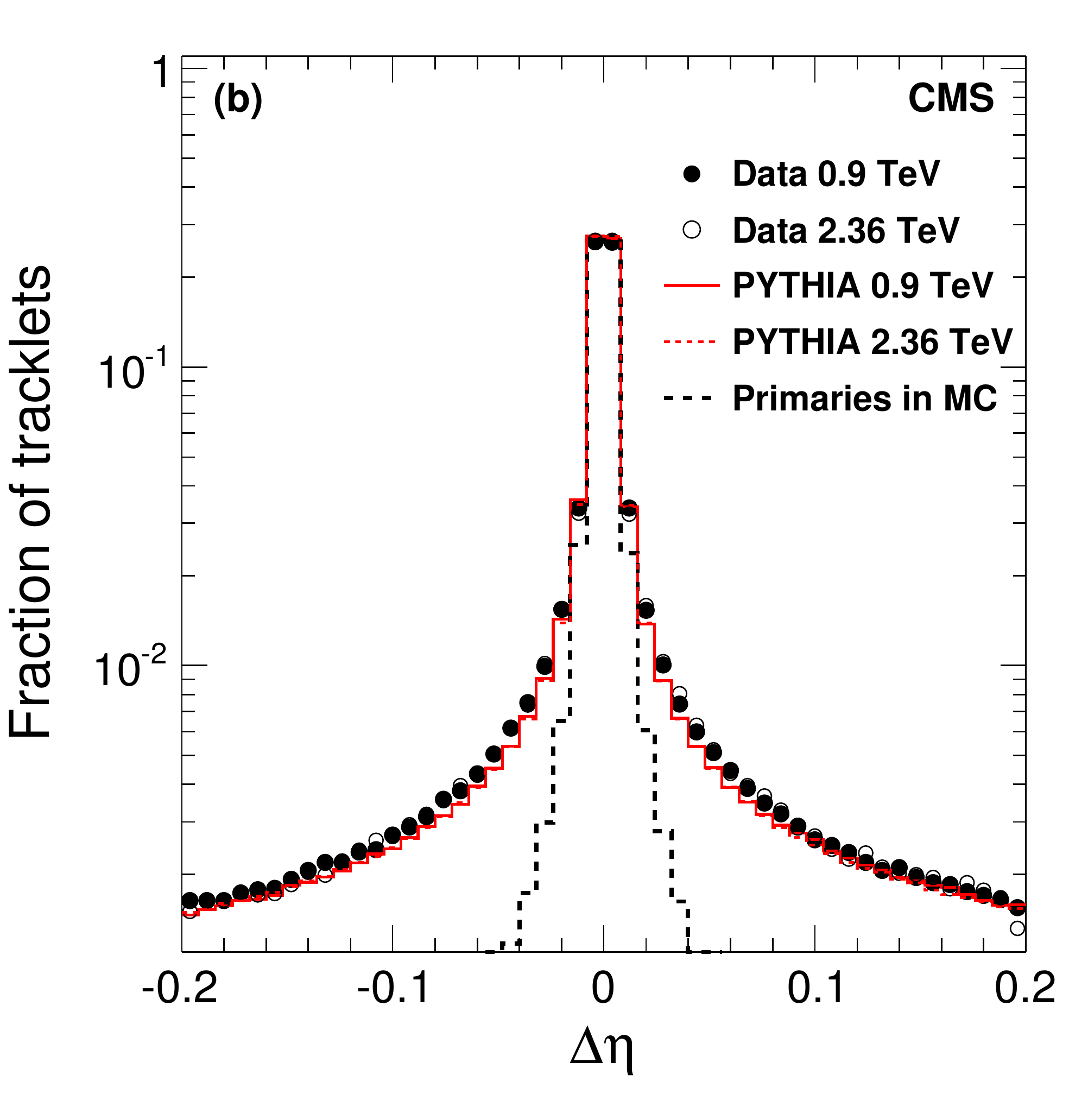}
}
\caption{
    \label{tracklet_recovz}
	\subref{background_1} Pixel cluster length along $z$ as a function of 
        $\eta$ for the 900 GeV data. The solid line illustrates the cut applied in the cluster 
	counting method.
	\subref{background_2} The $\deta$ distribution of clusters on 
	tracklets in the data (dots and circles), together with 
	the distribution obtained from the \PYTHIA simulation (solid 
	and dotted lines), for both 0.9 and 2.36~TeV 
	collision energy. The dashed line shows the $\deta$ 
	distribution of clusters for primary charged-particle tracks in 
	the Monte Carlo simulation at 0.9~TeV. The tail of the $\deta$ 
	distribution comes from the combinatorial background. 
}
\end{figure}

The cluster counting method correlates the observed pixel-cluster length in 
the $z$ direction, expressed in number of pixels, with the expected path 
length traveled by a primary particle at a given $\eta$ value. For 
primary particles the cluster length in $z$ is proportional to 
$|\cot\theta|$ (or $|\sinh\eta|$) as displayed in 
Fig.~\ref{tracklet_recovz}a. 
Small clusters at large $|\eta|$ are due to loopers, secondary particles 
and daughters of long-lived hadrons. Clusters from these background particles
were efficiently removed by the cluster-length cut 
represented by the solid line in Fig.~\ref{tracklet_recovz}a.
To allow for an efficient background rejection, only the barrel part of 
the pixel detector was used, where the detector units are parallel to the 
beam axis, as opposed to the pixel end-caps. Furthermore, the
$\eta$ range for the cluster counting was restricted to
$|\eta|<2$ to avoid acceptance problems due to the slightly off-centred
position of the luminous region.

The event selection efficiency and the SD contribution for a given total 
multiplicity of selected clusters ($M$) for each pixel barrel layer
can be determined from Monte Carlo simulation. The overall change of 
the \dnchdeta\ value due to this correction is 9\% for both 
collision energies.


The fraction of clusters created by loopers above the cluster-length cut 
(1\% and 5\% for $\eta=2$ and $\eta=0.5$, respectively) can be estimated 
by measuring the total number of clusters below the cut in data, corrected 
by the ratio of looper clusters below and above the cut in simulated events. 
The number of clusters eliminated by the cluster-length cut was found to 
be higher in data than in simulated events by 10-20\% due to a slightly 
larger abundance of secondary particles and loopers, while the 
observed number and length distributions of clusters above the cut 
was found to agree with the simulation.

The corrections for absorption in the beam pipe and detector material,
secondary particles, daughters of long-lived hadrons, delta-ray electrons 
and double hits caused by geometrically overlapping detector units were all 
evaluated, in bins of $\eta$ and $M$, with simulated data. The size of these 
corrections is 10\%, 23\% and 41\% for the first, second and third 
detector layer, respectively.
Varying the charged-particle multiplicity in the event generator by 50\% 
only causes a $\pm3\%$ relative change in these corrections. Their 
dependence on $\eta$ and pixel-cluster multiplicity is similarly small.

\subsection{Pixel-tracklet method}
\label{sect:reco_tracklets}

This method was first used to measure charged-hadron multiplicities by 
the PHOBOS experiment at RHIC~\cite{Back:2000gw}.
Pixel tracklets are constructed from combinations of two pixel hits in any 
two pixel barrel layers. The difference in the angular positions of the 
two clusters with respect to the PV, \deta\ and \dphi, are 
calculated for each tracklet. If two tracklets share a hit, the tracklet 
with the larger \deta\ is discarded. The \deta\ distribution of 
reconstructed tracklets is shown in Fig.~\ref{tracklet_recovz}b, 
together with the corresponding distribution from simulated data and a separate 
distribution for simulated primary particles only. Tracklets from primary
particles display a sharp peak at \deta\ $=0$, while the tracklets from the combinatorial 
background have an extended tail. The simulated \deta\ distributions are in good 
agreement with data.

To suppress the combinatorial background, only tracklets with 
$|\deta| < 0.1$ and $|\dphi| < 1.0$ were selected. Since the combinatorial 
background is flat in \dphi, the remaining fraction of background 
tracklets in the signal region $|\dphi| < 1.0$ can be estimated from 
tracklets with $1.0<|\dphi|<2.0$. This data-driven estimate of the background 
accurately describes the raw \deta\ distribution of tracklets for $|\deta| > 2$, 
where no signal from primary particles is expected from the MC simulation.
Typical values of this estimated background fraction in the signal region increase 
with $|\eta|$ from 2\% to 30\%. The $\eta$ range for the tracklet method 
was restricted to $|\eta|<2$ to avoid a large acceptance correction.


The contribution from secondary particles, reconstruction efficiency and 
geometrical acceptance needs to be accounted for to determine the number of 
primary charged hadrons. These correction factors were calculated using 
\PYTHIA\ simulations for background-subtracted tracklets in bins of 
$z$ position of the PV, pseudorapidity, and tracklet multiplicity. The 
magnitude of the correction varies with $|\eta|$ from 0 to 20\%. The correction
factors were also cross-checked by \PHOJET\ simulations and only cause a
2-3\% change in the \dnchdeta\ result.

The correction for the event selection efficiency and the SD contribution 
was determined for each tracklet multiplicity bin. 
The overall change in the \dnchdeta\ value 
due to this correction is about 8\% at $\eta=0$.

\subsection{Tracking method}
\label{sect:reco_tracks}

Pixel and SST detectors were used to reconstruct tracks, including both 
barrel and end-cap layers. The acceptance was limited to $|\eta|<2.4$ 
to avoid edge effects. The iterative reconstruction procedure described below follows 
Refs.~\cite{Cucciarelli:2006mt,Speer:2005dp}, but was further optimized 
for primary-track reconstruction in minimum bias events.

In the first step of track reconstruction, tracks with three pixel hits
(triplets) are built using the $x$ and $y$ positions of the beam spot 
and the $z$ coordinate of the primary vertex as constraints. These clean 
pixel tracks are used as seeds for the Kalman-filter-based trajectory-building 
algorithm in the SST. The resulting trajectories are stored.
Before the second tracking step, the pixel and strip hits associated with the
tracks found in the first step are removed from further consideration. 
The second step uses pixel triplet seeds as well, but does not require a 
vertex constraint and has a looser transverse impact parameter requirement 
than in the first step.
After removal of hits associated with tracks found in the second step, the 
third tracking step finds primary tracks seeded by two hits in the pixel 
detector. At least three hits were required for a track to be accepted.

Tracks found during the three iterative steps were collected and a second 
iteration of the PV reconstruction, as described in 
Section~\ref{sect:reco_vertex}, was performed to refine primary vertex 
position determination. Finally, the tracks were refit with the 
corresponding vertex constraint, thus improving their $\eta$ and \pt\ 
resolution.

In this analysis, a reconstructed track was considered as a 
primary-track candidate if it is compatible with originating from the PV 
($d_{\rm T} < \min(4\sigma_{\rm T},$~$\mathrm{0.2~cm})$ and 
$d_z < 4\sigma_z$, where $d_z$ is the distance between the point 
of the closest approach of the track to the beam axis and the PV 
along the $z$ direction). 

Studies with simulated events showed that the combined 
geometrical acceptance and reconstruction efficiency for the  
tracking method exceeds 50\% around \pt\ $\approx$ 0.1, 0.2 
and 0.3~\GeVc\ for pions, kaons and protons, respectively. The efficiency
is about 96\% in the $|\eta|<1$ region for $\pt >$ 0.25~\GeVc, and is 
above 80\% for pions at $\pt =$ 0.15~\GeVc. By requiring the 
geometrical shapes of the pixel clusters to be consistent with the crossing 
angle and direction of the track, the fraction of fake tracks was kept 
below 1\%. The fraction of duplicated tracks (e.g., from loopers) was 
estimated to be about 0.1\% in the central region, rising to 0.5\% at 
large $|\eta|$.


The measured yield in data was corrected, based on MC simulation 
and comparisons with data, for geometrical acceptance
(2\% correction for $p_T>200$~MeV/c), efficiency of the reconstruction 
algorithm (5-10\% for $p_T>300$~MeV/c), fake and duplicate 
tracks ($<$1\% each). The contamination of less than $2\%$ 
from decay products of long-lived hadrons, photon conversions and inelastic hadronic 
interactions with the detector material was also subtracted.
To obtain the \dnchdeta\ result from the 
\pt spectrum, an extrapolation to $\pt=0$ was necessary, resulting in 
an increase of 5\% in the estimated number of charged hadrons.

\begin{figure}[tp]
\centering
\subfigure{
        \label{vz_npix_2}
    \includegraphics[width=0.45\textwidth]{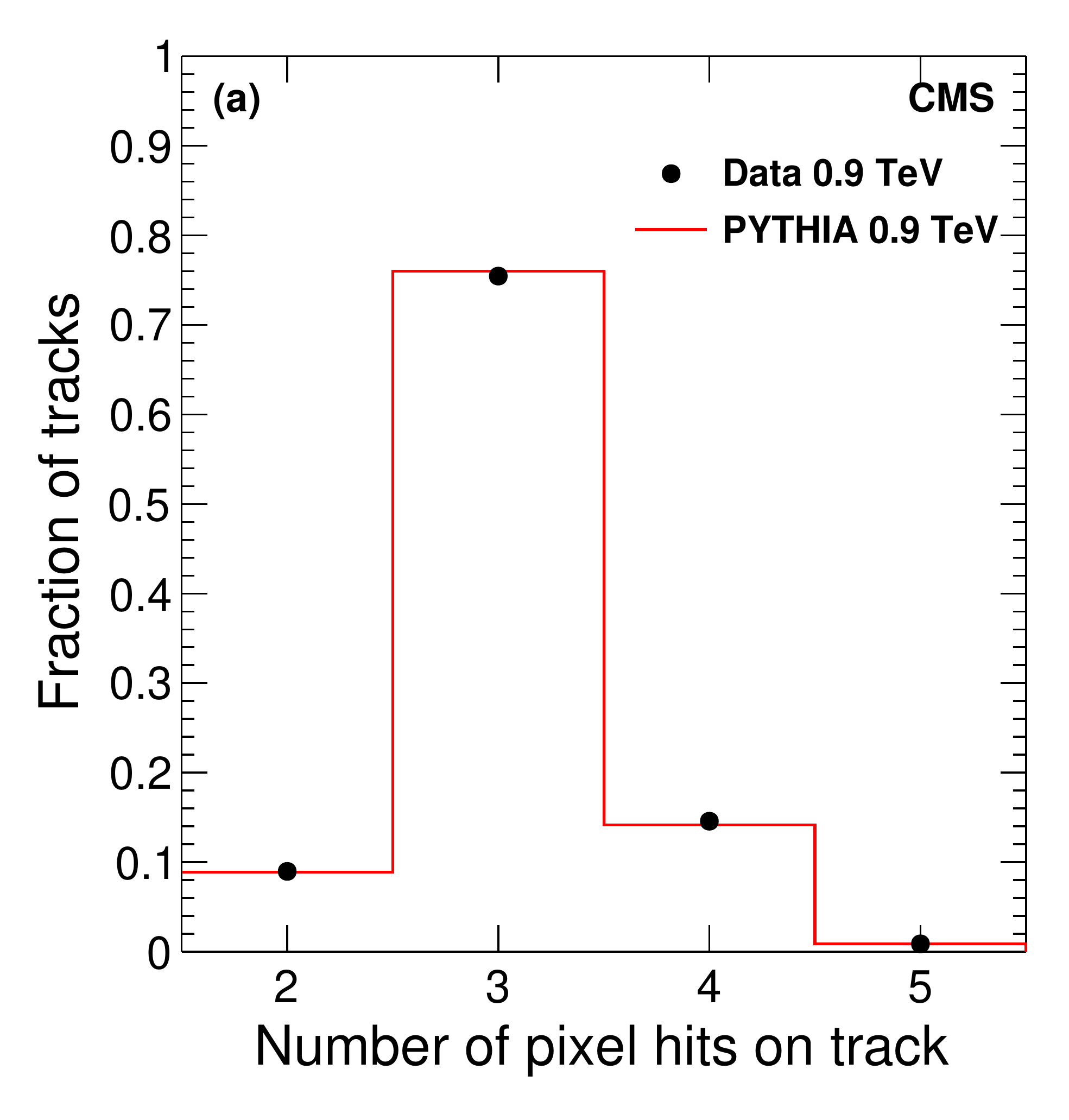}}
\hspace{1cm}
\subfigure{
        \label{nstrip}
    \includegraphics[width=0.45\textwidth]{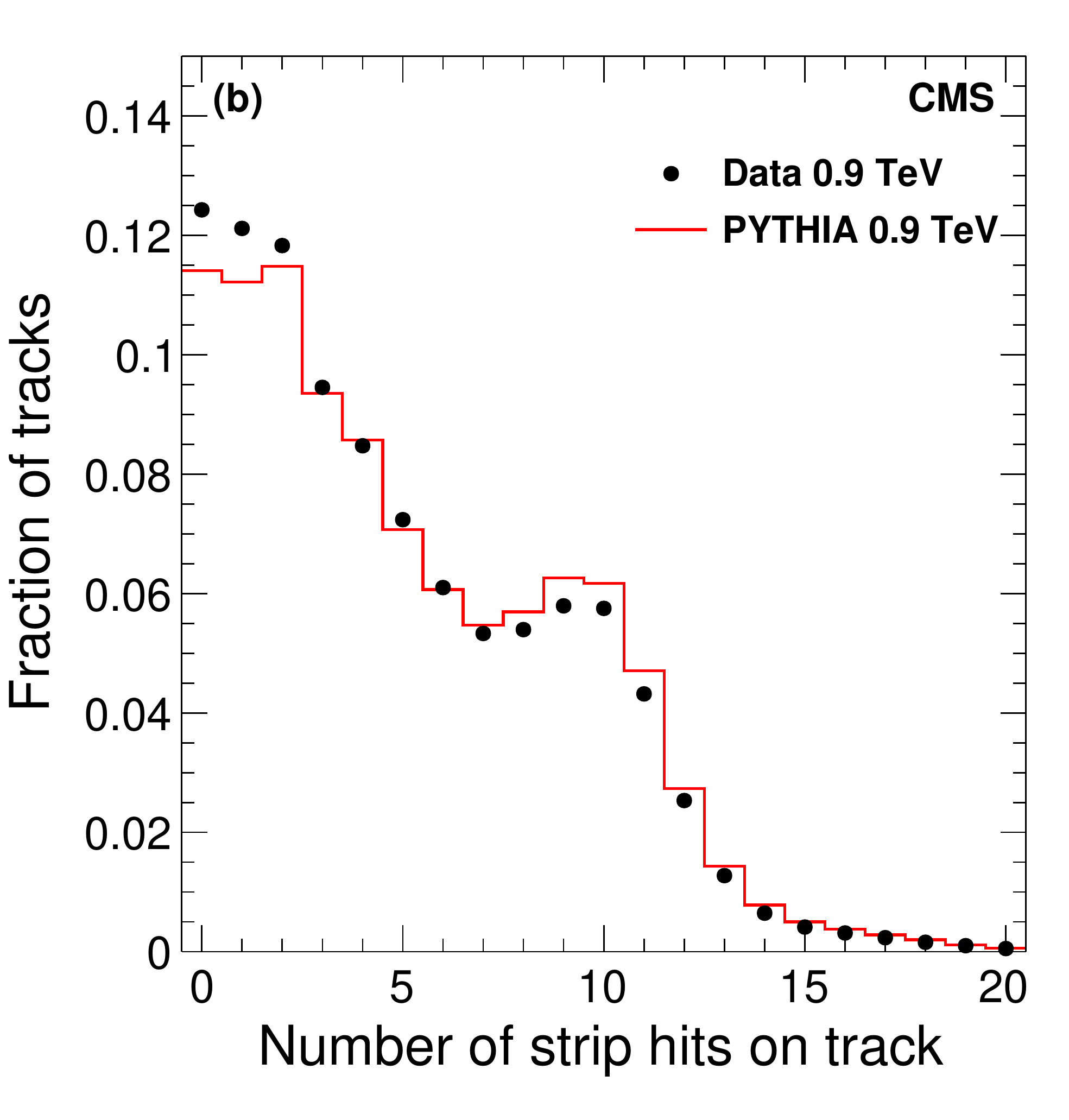}}
\caption{
    \label{vz_npix}
        \subref{vz_npix_2} The distribution of the number of pixel hits
        attached to reconstructed tracks in the
        region of $|\eta|<1$ (closed circles), compared to the
        CMS detector simulation (histogram).
        \subref{nstrip} The distribution of the number of hits in the SST detector
        attached to reconstructed tracks in the
        region of $|\eta|<1$ (closed circles), compared to
        the CMS detector simulation (histogram). }
\end{figure}

Corrections based on the average hit efficiency of pixel 
layers, size of the beam spot, longitudinal and transverse impact-parameter resolutions of pixel tracks were validated with data. As an
example, the average number of pixel and strip hits found on tracks in 
the range $|\eta|<1$ is shown in Figs.~\ref{vz_npix}a and \ref{vz_npix}b, 
together with the expectation from \PYTHIA.
Somewhat fewer particles are predicted with $p_T<500$~MeV/c than 
seen in the data, which results in the small difference in the number of 
tracks with few SST hits in Fig.~\ref{vz_npix}b. This small 
difference, which originates from limitations of the \PYTHIA 
generator, does not affect the final measurement.
The correction for the event selection efficiency and the SD contribution 
was determined for each track multiplicity bin, and has an overall 
magnitude of 8.3\%.

\section{Results}
\label{sect:results}

\subsection {Systematic uncertainties}
\label{sub:syserr}

Various corrections and their event-selection and model dependence 
contribute to the systematic uncertainties of the measured quantities. A 
summary of these systematic uncertainties averaged over $\eta$ and \pt\ is 
given in Table~\ref{tab:syst-table} and discussed below.

The uncertainties related to the trigger bias and to the 
event selection are common to all the analysis methods. 
The efficiency of the trigger and event-selection was corrected for 
by the prediction of the \PYTHIA D6T event generator combined with the 
full \GEANTfour\ simulation of the CMS detector. The material description 
relevant for this analysis was verified by studies of photon conversion 
probabilities in the data, found to be in agreement with those obtained 
from the simulation.

Because single- and double-diffractive \pp\ collisions 
have much smaller charged-hadron multiplicities per 
event than non-diffractive events, they contribute to the uncertainty in 
the measured \dnchdeta\ mostly through the uncertainty in the fraction of 
SD and DD events passing the event selection criteria.
The fractions of SD events for $\roots =$~0.9~TeV in \PYTHIA and 
\PHOJET (Table~\ref{tab:difffrac}) are 23\% and 19\%,
respectively. The UA5 experiment measured 
15.5\% for this fraction~\cite{Ansorge:1986xq}. Based on the 
simulated trigger efficiencies for the different event types,
only 5.5\% of events passing the analysis event selection 
are expected to be single-diffractive events. From the aforementioned 
variations of SD fractions, an uncertainty of $\pm$1\% is attributed to this 
correction. The contribution of the uncertainty of the fraction of DD 
events was estimated similarly to be $\pm$1\%. Since underestimated 
DD and SD fractions both lead to an underestimated \dnchdeta\ 
result, a conservative linear sum of 2\% was assigned to the 
above systematic uncertainty. The trigger efficiency of the BSC 
is more than 98\% for events with a valid vertex, and even a 5\% 
uncertainty in single-particle detection efficiency 
of its individual segments results in a negligible uncertainty in 
the final result.
The trigger efficiency of the BSC and the event selection 
efficiency of the HF detector were both measured from
data and found to be consistent within 1\% with the MC simulation.
The total systematic uncertainty from propagating all event selection and 
trigger related uncertainties is 3\%. The measurement of the average 
transverse momentum is less sensitive to the trigger selection efficiency. 
A smaller, 1\% uncertainty was therefore assigned to that result.

The geometrical acceptance was studied by comparing the hit occupancy 
of the pixel barrel with the predictions from the simulation. 
The efficiency of the pixel hit reconstruction was estimated using tracks 
propagated from the SST to the pixel detector and by extrapolating 
pixel tracklets to the unused pixel barrel layer. 
The measured pixel hit efficiency was found to exceed 99\% with a 0.5\% 
uncertainty from both methods, which propagates into 0.5\% uncertainty in the 
pixel-counting-based, 1\% in the tracklet-based, and 0.3\% in the 
track-based results. If the collected charge in one or more
pixels in a cluster remains below the threshold, the cluster may be split.
The splitting rate was estimated from the geometrical distance distributions 
of close-by pixel clusters found in the data and in the Monte Carlo simulation 
and found to be 0.5-0.9\% in the simulation and 1.0-1.5\% in data.

The uncertainty related to the cluster and tracklet selections was 
estimated by varying the selection cuts. 
An additional 3\% and 2\% uncertainty was assigned to the tracklet and 
track reconstruction algorithm efficiencies, 
respectively. Corrections for loopers and secondary particles are 
simulation dependent; the tracklet- and pixel-counting-based methods have 
low rejection power compared to the tracking method, thus carry a 
larger systematic uncertainty (as shown in Table~\ref{tab:syst-table}). 

The effects of the geometrical misalignment of the pixel barrel detector 
were simulated and a 1\% uncertainty was assigned to the results from the 
tracklet-based method. Hits from beam-induced backgrounds 
coinciding with the collision were estimated to be very rare, and
a conservative 1\% random hit contribution was propagated to obtain the 
uncertainty of the results. The corrections for multiple track counting 
and fake track rate were estimated 
from the Monte Carlo simulation and found to be less than 1\%.
The uncertainty of the extrapolation to the full $p_T$-range depends on the low-$p_T$
reach of the three methods and varies between 0.2 and 0.5\%. While the 
sources of uncertainties are largely independent from each other, they are 
correlated among all the data points.

\begin{table}[t]
\caption{\label{table:systematics}Summary of systematic uncertainties.
While the various sources of uncertainties are largely independent,
most of the uncertainties are correlated between data points and between the
analysis methods. The event selection and acceptance uncertainty is common 
to the three methods and affects them in the same way. The values in 
parentheses apply to the $\langle p_T \rangle$ measurement.}
\label{tab:syst-table}
\begin{center}
\begin{tabular}{lcccc}
\hline
\hline
 Source                                          & Pixel Counting [\%] & Tracklet [\%] & Tracking [\%] \\
\hline
 Correction on event selection                   & 3.0           & 3.0		       & 3.0 (1.0)      \\
 Acceptance uncertainty                          & 1.0           & 1.0                 & 1.0       \\
 Pixel hit efficiency                            & 0.5           & 1.0                 & 0.3       \\
 Pixel cluster splitting                         & 1.0           & 0.4                 & 0.2       \\
 Tracklet and cluster selection                  & 3.0           & 0.5                 & -         \\
 Efficiency of the reconstruction	 	 		 & -             & 3.0                 & 2.0       \\
 Correction of looper hits                       & 2.0           & 1.0                 & -         \\
 Correction of secondary particles               & 2.0           & 1.0                 & 1.0       \\
 Misalignment, different scenarios               & -             & 1.0                 & 0.1       \\
 Random hits from beam halo                      & 1.0           & 0.2                 & 0.1       \\
 Multiple track counting                         & -             & -                   & 0.1       \\
 Fake track rate                                 & -             & -                   & 0.5       \\
 \pt\ extrapolation                              & 0.2           & 0.3                 & 0.5       \\
\hline
 Total, excl. common uncertainties               & 4.4           & 3.7                 & 2.4       \\
 Total, incl. common uncert. of 3.2\%            & 5.4           & 4.9                 & 4.0 (2.8) \\
\hline
\hline
\end{tabular}
\end{center}
\end{table}

\subsection{Charged hadron transverse-momentum distributions}

\begin{figure}
 \begin{flushleft}
  \subfigure{\label{fig:spectra_a}
  \includegraphics[width=0.48\textwidth,height=0.5\textwidth]{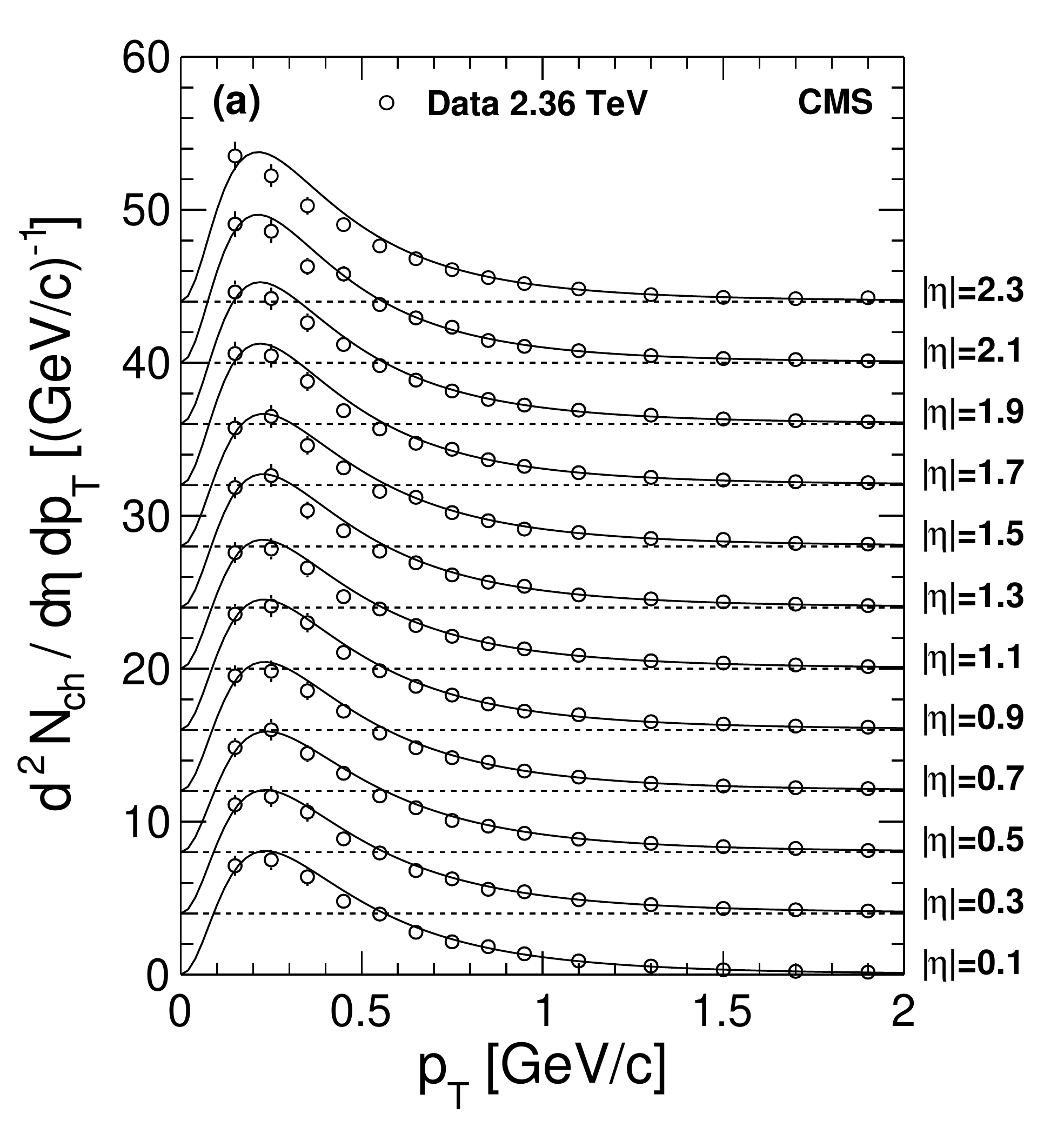}}
  \hspace{0cm}
  \subfigure{\label{fig:spectra_b}
  \includegraphics[width=0.48\textwidth,height=0.5\textwidth]{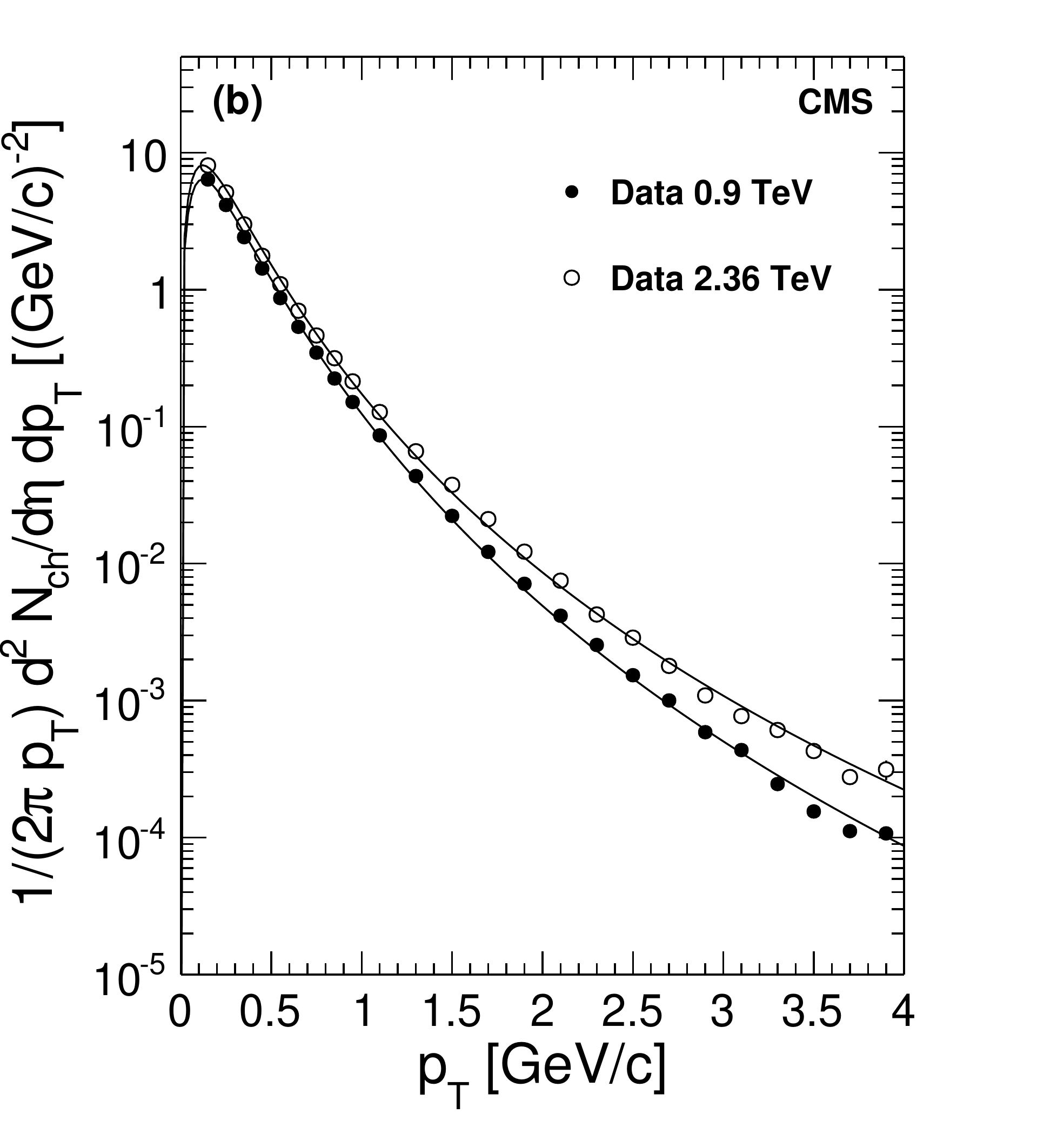}}
 \end{flushleft}
 \caption{(a) Measured differential yield of charged hadrons in the
range $|\eta| < 2.4$ in $0.2$-unit-wide bins of $|\eta|$ for the 2.36~TeV 
data. The measured values with systematic uncertainties (symbols) and the 
fit functions (Eq.~\ref{eq:tsallis}) are displayed. The values with 
increasing $\eta$ are successively shifted by four units along the 
vertical axis.
(b) Measured yield of charged hadrons for $|\eta| < 2.4$ with systematic 
uncertainties (symbols), fit with the empirical function 
(Eq.~\ref{eq:tsallis}).
}
 \label{fig:spectra}
\end{figure}

Tracks with $|\eta|<2.4$ and $\pt>0.1$~\GeVc\ were used for the
measurement of \dnchdpt. The measured average charged-hadron 
yields per NSD event are shown in Fig.~\ref{fig:spectra}a, as a function 
of \pt\ in bins of $|\eta|$. The yields were fit by the Tsallis function 
(Eq.~\ref{eq:tsallis}), which empirically describes both the low-\pt\ 
exponential and the high-\pt\ power-law behaviours 
\cite{Tsallis:1987eu,Wilk:2008ue}:

\begin{equation}
 E \frac{d^3N_{\mathrm{ch}}}{d p^3} =
\frac{1}{2\pi p_T} \frac{E}{p} \frac{d^2N_{\mathrm{ch}}}{d\eta dp_T} =
C(n,T,m) \frac{dN_{\mathrm{ch}}}{dy}\left(1 + \frac{E_T}{nT}\right)^{-n} ,
 \label{eq:tsallis}
\end{equation}

\noindent where $y=0.5\ln[(E+p_z)/(E-p_z)]$ is the rapidity;
$C(n,T,m)$ is a normalization constant that depends on $n$, $T$ and $m$;
$\ET = \sqrt{m^2 + \pt^2} - m$ and $m$ is the charged 
pion mass. This function provides both the inverse slope parameter $T$, 
characteristic for low \pt, and the exponent $n$, which parameterizes the 
high-\pt\ power-law tail. These fit parameters change by less than 5\% 
with $\eta$, thus a fit to the whole region $|\eta|<2.4$ was performed. 
The \pt\ spectrum of charged hadrons,
$1/(2\pi \pt)d^2N_{\rm ch}/d\eta d \pt$, in the region $|\eta| < 2.4$, 
was also fit with the empirical function (Eq.~\ref{eq:tsallis}) and is 
shown in Fig.~\ref{fig:spectra}b.
The \pt resolution of the CMS tracker was found to have a negligible 
effect on the measured spectral shape and was therefore ignored in the fit function. 
For the 0.9~TeV data, the inverse slope parameter and the exponent 
were found to be $T = 0.13 \pm 0.01$~\GeV and $n = 7.7 \pm 0.2$. 
For the 2.36~TeV data, the values were $T = 0.14 \pm 0.01$~\GeV and 
$n = 6.7 \pm 0.2$.
The average transverse momentum, calculated from the measured data points 
adding the low- and high-\pt\ extrapolations from the fit is 
$\langle \pt \rangle = 0.46 \pm 0.01$~(stat.)~$\pm$~0.01~(syst.)~\GeVc 
for the 0.9~TeV and 
$0.50 \pm 0.01$~(stat.)~$\pm$~0.01~(syst.)~\GeVc for the 2.36~TeV data. 

The \dnchdeta\ spectrum was obtained by summing the measured differential 
yields for $0.1 < \pt < 3.5$~\GeVc\ and adding the result to the integral 
of the fit function for $\pt < 0.1$~\GeVc and $\pt > 3.5$~\GeVc.
The latter term amounts to 5\% of the total.

\subsection{Charged hadron pseudorapidity density}

The summary of results on the pseudorapidity density distribution of 
charged hadrons is shown in Fig.~\ref{dndeta_results}.
The \dnchdeta\ results for the three layers in the cluster-counting method 
and the three layer-pairs in the pixel-tracklet method are consistent 
within 3\%. These results from the various layers and from the different 
layer pairs were combined to provide one set of data from each analysis
method, as shown in Fig.~\ref{dndeta_results}a.
The error bars include the systematic uncertainties of about 
2.4--4.4\% specific to each method, estimated from the variations of 
model parameters in the simulation used for corrections and the 
uncertainties in the data-driven corrections. The systematic uncertainties 
common to all the three methods, which amount to 3.2\%, are not shown.
The results from the three analysis methods are in agreement.
The larger fraction of background hits in the data 
compared to simulation affects the cluster-counting method differently 
from the other two, which results in a small difference at high $\eta$, 
well accounted for by the systematic uncertainty of the measurement.

\section{Discussion}
\label{sect:discuss}

\begin{figure}[h]
 \begin{center}
  \subfigure{\label{roots_dependence_a}
  \includegraphics[width=0.48\textwidth]{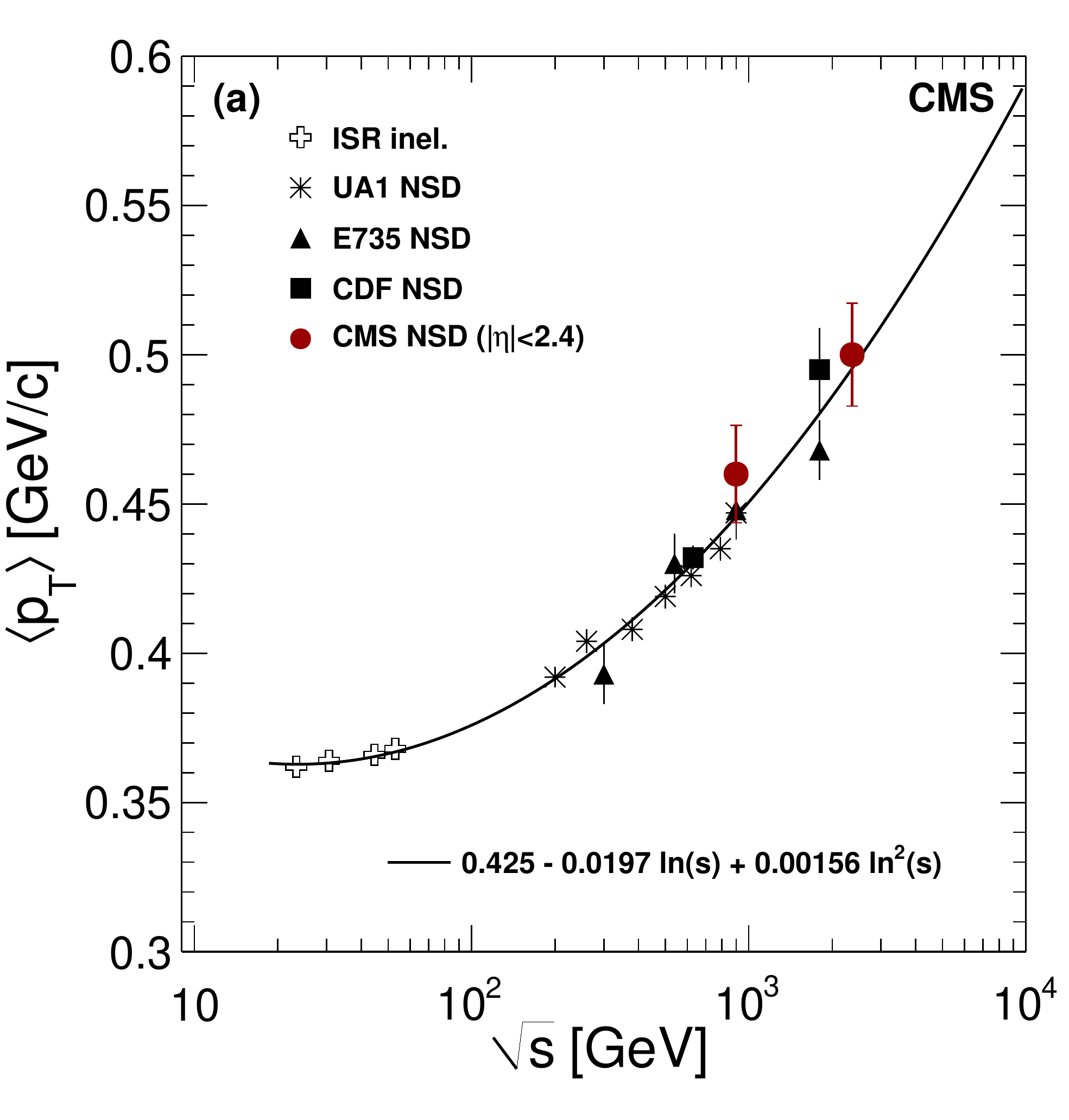}}
  \hspace{0cm}
  \subfigure{\label{roots_dependence_b}
  \includegraphics[width=0.48\textwidth]{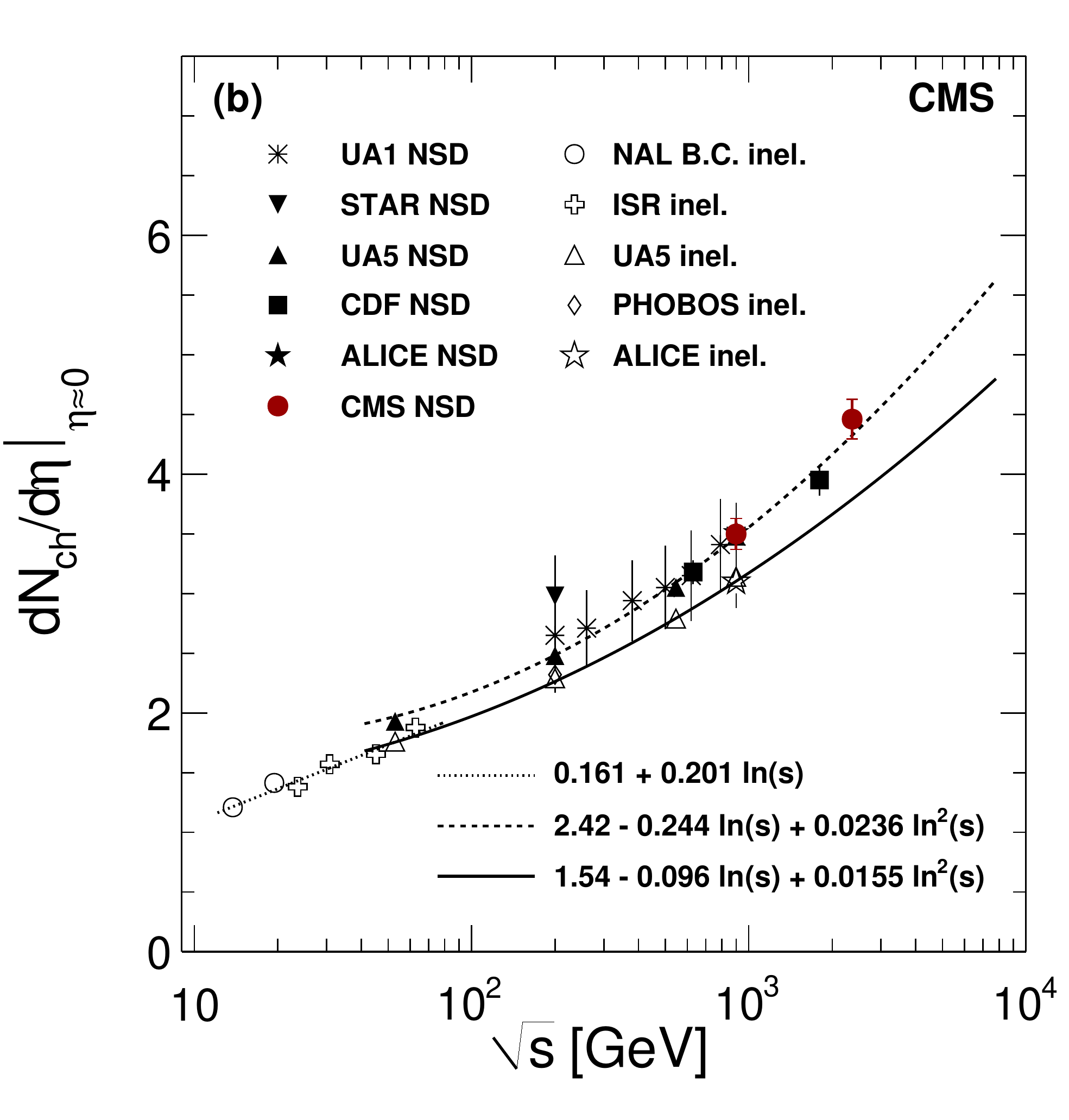}}
 \end{center}
 \caption{
(a) Energy dependence of the average transverse momentum of
charged hadrons. The CMS data points are evaluated for the range
$|\eta|<2.4$. Data of other experiments are taken from Refs.
\cite{Rossi:1974if,Albajar:1989an,Alexopoulos:1988na,Abe:1988yu}.
The curve shows the fit to the data points of the form
$\langle \pt \rangle=0.425-0.0197\ln(s)+0.00156\ln^2(s)$ with 
$\langle \pt \rangle$ in GeV/c and $s$ in GeV$^2$. The error bars on 
the CMS data points include systematic uncertainties. 
(b) Charged-hadron pseudorapidity density in the central region as a 
function of centre-of-mass energy in \pp\ and \pbarp\ collisions including 
lower energy data from Refs.
\cite{Whitmore:1973ri,Thome:1977ky,Alner:1986xu,Albajar:1989an,Abe:1989td,alice,
star:2008ez,Nouicer:2004ke}, together with various empirical parameterizations
fit to the data corresponding to the inelastic (solid and dotted 
curves with open symbols) and to the NSD (dashed curve with solid symbols) 
event selection. The error bars indicate systematic uncertainties, when 
available.}
 \label{roots_dependence}
\end{figure}

The average transverse-momentum and pseudorapidity densities of charged 
hadrons derived from the measured data can be compared to results from earlier 
experiments as a function of the collision energy.
The average transverse momentum of charged hadrons was obtained from 
the fits (Eq.~\ref{eq:tsallis}) to the transverse-momentum 
spectrum (Fig.~\ref{fig:spectra}b). At low energies the energy 
dependence of $\langle \pt \rangle$ can be described by a quadratic 
function of $\ln{s}$. The $\langle \pt \rangle$ from this 
measurement, shown in Fig.~\ref{roots_dependence}a, follows 
the general trend. At 0.9~TeV it is similar to the results from 
\ppbar\ collisions at the same energy~\cite{Albajar:1989an}.

\begin{figure}[tp]
  \begin{center}
    \subfigure{\label{dndeta_results_a}
    \includegraphics[width=0.48\textwidth]{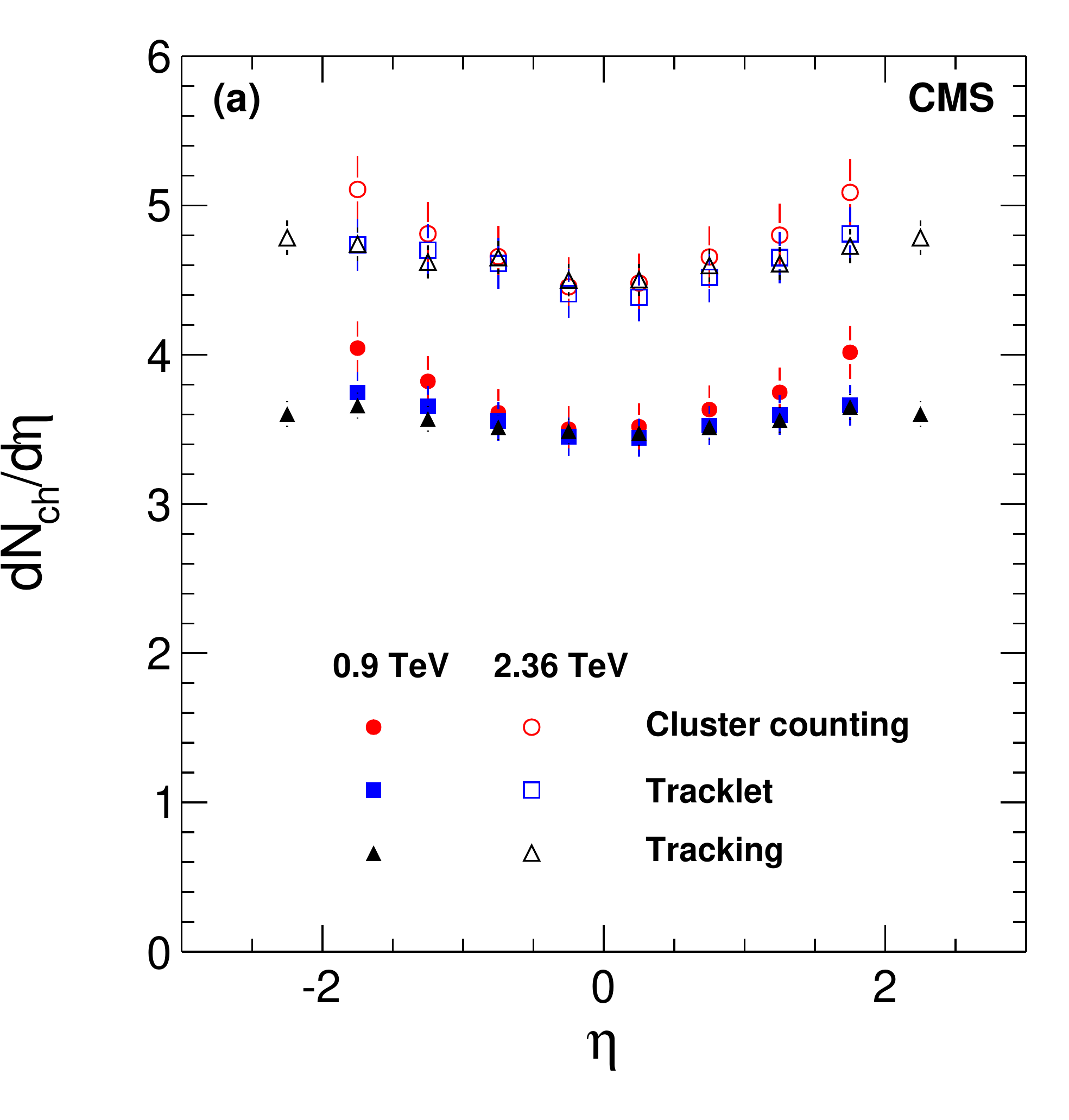}}
    \hspace{0cm}
    \subfigure{\label{dndeta_results_b}
    \includegraphics[width=0.48\textwidth]{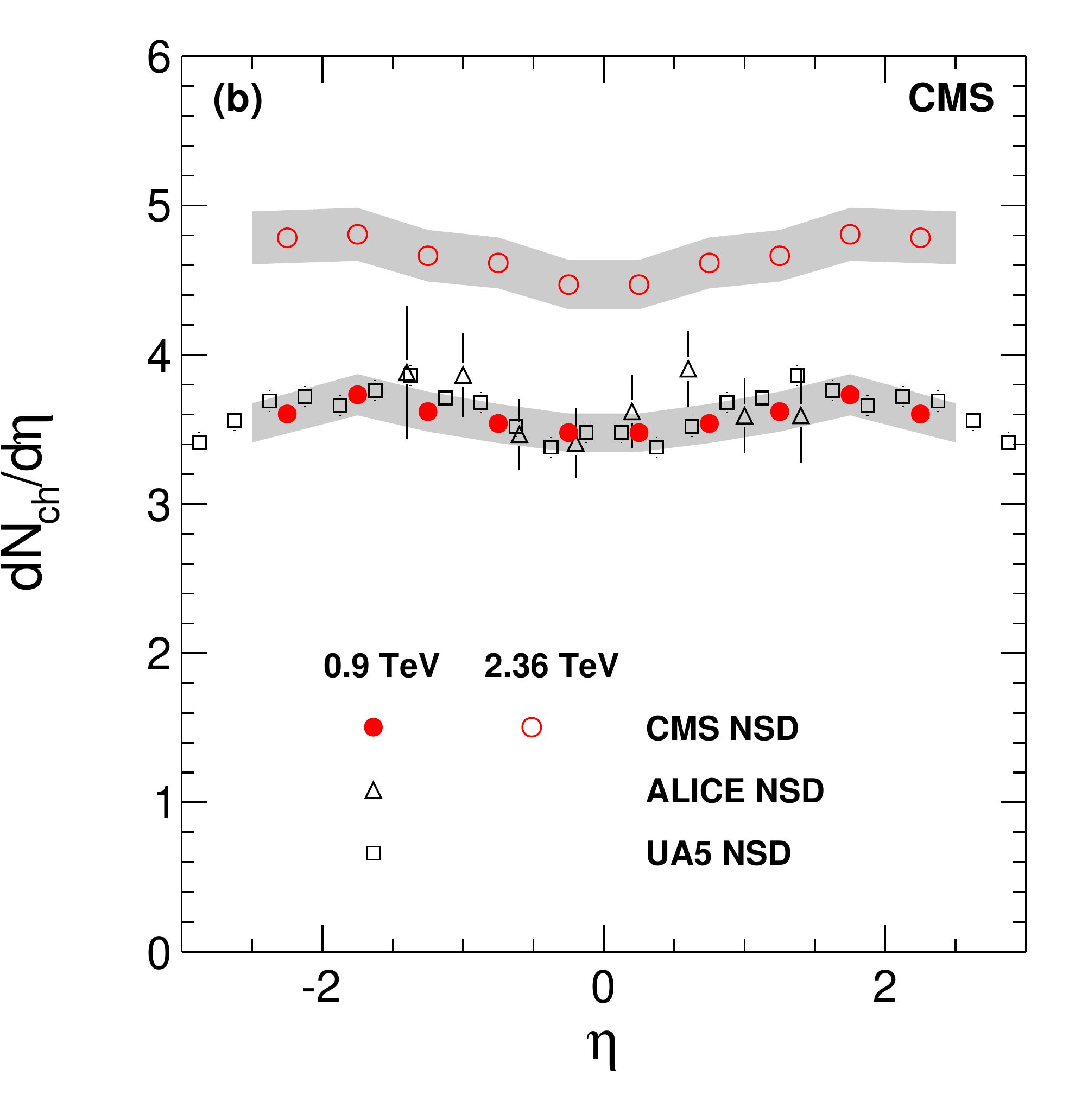}}
    \caption{(a) Reconstructed \dnchdeta\ distributions obtained from the 
cluster counting (dots with error bars), tracklet (squares) and tracking 
(triangles) methods, in \pp\ collisions at 0.9~TeV (filled symbols) and 
2.36~TeV (open symbols).
The error bars include systematic uncertainties (as discussed in the 
text), excluding those common to all the methods.
(b) Reconstructed \dnchdeta\ distributions averaged over
the cluster counting, tracklet and tracking methods
(circles), compared to data from the UA5~\cite{Alner:1986xu} 
(open squares) and from the ALICE \cite{alice} (open triangles) 
experiments at 0.9~TeV, and the averaged result over the three methods at 
2.36~TeV (open circles). The CMS and UA5 data points are symmetrized 
in $\eta$. The shaded band represents systematic 
uncertainties of this measurement, which are largely correlated 
point-to-point. The error bars on the UA5 and ALICE data 
points are statistical only.}
    \label{dndeta_results}
  \end{center}
\end{figure}

The \dnchdeta\ distribution was calculated as the weighted average of the 
data from the three reconstruction methods, taking into account their 
systematic uncertainties, excluding the common ones, as listed in 
Table~\ref{tab:syst-table}. The averaged result is shown in 
Fig.~\ref{dndeta_results}b and compared to measurements at the same 
accelerator (ALICE, \pp\,\cite{alice}) and to previous measurements at the 
same energy but with different colliding particles 
(UA5, \pbarp\,\cite{Alner:1986xu}). The shaded error band on the CMS data 
indicates systematic uncertainties, while the error bars on the data from 
UA5 and ALICE display statistical uncertainties only.
No significant difference is observed between 
the \dnchdeta\ distributions measured in \pp\ and \pbarp\ collisions at 
$\sqrt{s}=0.9$~TeV.

The \dnchdeta\ distribution is weakly $\eta$-dependent, with a slow 
increase towards higher $\eta$ values, and an indication of a decrease 
at $|\eta|>2$ for the 0.9~TeV data. At 2.36~TeV, the entire distribution 
is wider due to the increased collision energy hence the larger 
$\eta$ range available for inclusive particle production.
For $|\eta| < 0.5$, the corrected results average to 
$\dnchdeta = 3.48 \pm 0.02$~(stat.)~$\pm$~0.13~(syst.) and
$\dnchdeta = 4.47 \pm 0.04$~(stat.)~$\pm$~0.16~(syst.) for 
NSD events at $\roots = 0.9$ and 2.36 TeV. 
The increase of $(28.4 \pm 1.4 \pm 2.6)\%$ from 0.9 to 2.36~TeV is 
significantly larger than the 18.5\% (14.5\%) increase predicted by the
\PYTHIA (\PHOJET) model tunes used in this analysis.
The collision energy dependence of the 
measured $\dnchdeta|_{\eta \approx 0}$ is shown in 
Fig.~\ref{roots_dependence}b, 
which includes data from the NAL Bubble 
Chamber~\cite{Whitmore:1973ri}, the
ISR~\cite{Thome:1977ky}, and UA1~\cite{Albajar:1989an},
UA5~\cite{Alner:1986xu}, CDF~\cite{Abe:1989td},
STAR~\cite{star:2008ez}, PHOBOS~\cite{Nouicer:2004ke} and 
ALICE~\cite{alice}. The \dnchdeta\ measurement reported here is 
consistent with the previously observed trend.

\section{Summary}
\label{sect:summary}
Inclusive measurements of charged-hadron densities, \dnchdpt\, 
and \dnchdeta, have been presented based on the first \pp\ collisions 
recorded at $\roots = 0.9$ and 2.36~TeV by the CMS experiment during 
LHC commissioning in December 2009. The numerical values of the data 
presented in this paper can be found in Ref.~\cite{datatables}.
For NSD interactions, the average charged-hadron transverse momentum has 
been measured to be 
$0.46 \pm 0.01$~(stat.)~$\pm$~0.01~(syst.)~\GeVc at 0.9~TeV and 
$0.50 \pm 0.01$~(stat.)~$\pm$~0.01~(syst.)~\GeVc at 2.36~TeV. 
The three reconstruction methods employed for the \dnchdeta\ measurement 
have yielded consistent results, demonstrating the excellent performance 
and detailed understanding of the CMS tracker. The 
pseudorapidity density in the central region, 
$\dnchdeta |_{|\eta| < 0.5}$, has been measured to be
$3.48 \pm 0.02$~(stat.)~$\pm$~0.13~(syst.) at 0.9~TeV and
$4.47 \pm 0.04$~(stat.)~$\pm$~0.16~(syst.) at 2.36~TeV.
The results at 0.9~TeV have been found to be in agreement with 
previous measurements in \pbarp\ and \pp\ collisions. With the new
measurements at 2.36~TeV, which show a steeper-than-expected increase of 
charged-hadron multiplicity density with collision energy, the 
study of particle production in \pp\ collisions has been extended into a new 
energy regime.

\section*{Acknowledgements}

\label{sect:acknowledge}

We congratulate and express our gratitude to our colleagues in the CERN 
accelerator departments for the excellent performance of the LHC. We thank 
the technical and administrative staff at CERN and other CMS Institutes, 
and acknowledge support from: FMSR (Austria); FNRS and FWO (Belgium); 
CNPq, CAPES, FAPERJ, and FAPESP (Brazil); MES (Bulgaria); CERN; CAS, 
MoST, and NSFC (China); COLCIENCIAS (Colombia); MSES (Croatia); RPF 
(Cyprus); Academy of Sciences and NICPB (Estonia); Academy of Finland, 
ME, and HIP (Finland); CEA and CNRS/IN2P3 (France); BMBF, DFG, and HGF 
(Germany); GSRT (Greece); OTKA and NKTH (Hungary); DAE and DST (India); 
IPM (Iran); SFI (Ireland); INFN (Italy); NRF (Korea); LAS (Lithuania); 
CINVESTAV, CONACYT, SEP, and UASLP-FAI (Mexico); PAEC (Pakistan); SCSR 
(Poland); FCT (Portugal); JINR (Armenia, Belarus, Georgia, Ukraine, 
Uzbekistan); MST and MAE (Russia); MSTDS (Serbia); MICINN and CPAN 
(Spain); Swiss Funding Agencies (Switzerland); NSC (Taipei); TUBITAK and 
TAEK (Turkey); STFC (United Kingdom); DOE and NSF (USA). Individuals have 
received support from the Marie-Curie IEF program (European Union); the 
Leventis Foundation; the A. P. Sloan Foundation; and the Alexander von 
Humboldt Foundation.

\bibliography{auto_generated}   
\cleardoublepage\appendix\section{The CMS Collaboration \label{app:collab}}\begin{sloppypar}\hyphenpenalty=5000\textbf{Yerevan Physics Institute,  Yerevan,  Armenia}\\*[0pt]
V.~Khachatryan, A.M.~Sirunyan, A.~Tumasyan
\vskip\cmsinstskip
\textbf{Institut f\"{u}r Hochenergiephysik der OeAW,  Wien,  Austria}\\*[0pt]
W.~Adam, T.~Bergauer, M.~Dragicevic, J.~Er\"{o}, M.~Friedl, R.~Fr\"{u}hwirth, V.M.~Ghete, J.~Hammer\cmsAuthorMark{1}, S.~H\"{a}nsel, M.~Hoch, N.~H\"{o}rmann, J.~Hrubec, M.~Jeitler, G.~Kasieczka, M.~Krammer, D.~Liko, I.~Mikulec, M.~Pernicka, H.~Rohringer, R.~Sch\"{o}fbeck, J.~Strauss, A.~Taurok, F.~Teischinger, W.~Waltenberger, G.~Walzel, E.~Widl, C.-E.~Wulz
% ...
\vskip\cmsinstskip
% ...
1:~~Also at CERN, European Organization for Nuclear Research, Geneva, Switzerland\\
% ...

\end{sloppypar}
\end{document}